\documentclass{aa}

\usepackage[utf8]{inputenc}
\usepackage{graphicx}
\usepackage{amssymb}
\usepackage{amsmath}
\usepackage{float}
\usepackage{natbib}
\bibpunct{(}{)}{;}{a}{}{,} 
\usepackage{array}
\usepackage{hyperref}
\usepackage{subfig}
\usepackage[varg]{txfonts}
\usepackage{multirow}
\usepackage{siunitx}
    \DeclareSIUnit[]{\px}{px} 
    \DeclareSIUnit[]{\a}{a}    
    \DeclareSIUnit[]{\yr}{yr}    
    \DeclareSIUnit[]{\MEarth}{\textit{M}_{\rm E}} 
    \DeclareSIUnit[]{\MSun}{\textit{M}_{\rm \odot}}
\usepackage{upgreek}
\usepackage{textgreek}

\usepackage{mathabx}

\usepackage{titlesec}

\usepackage{xcolor}

\hypersetup{colorlinks=true,citecolor=blue}

\titleformat{\paragraph}
{\normalfont\normalsize\bfseries}{\theparagraph}{1em}{}
\titlespacing*{\paragraph}
{0pt}{3.25ex plus 1ex minus .2ex}{1.5ex plus .2ex}

\def\app#1#2{%
  \mathrel{%
    \setbox0=\hbox{$#1\sim$}%
    \setbox2=\hbox{%
      \rlap{\hbox{$#1\propto$}}%
      \lower1.1\ht0\box0%
    }%
    \raise0.25\ht2\box2%
  }%
}

\newcommand{\commentaire}[1]{}
\newcommand{\citealias}{\citetalias}

\DeclareTextSymbol{\degre}{T1}{6}
\DeclareTextSymbol{\degre}{OT1}{23}

\defcitealias{LP07}{LP07}          
\defcitealias{RNS14}{RNS14}    
\defcitealias{PE83}{PE83}         
\defcitealias{Guyonnet+2015}{GAARD}         
\defcitealias{CGPM-17.1}{BIPM 1983}
\defcitealias{CGPM-25.1}{BIPM 2014}
\defcitealias{CIPM-106}{BIPM 2017}

\defcitealias{Diaz+2016}{D16}
\defcitealias{Dumusque+2015}{D15}
\defcitealias{FTJ17}{F17a}
\defcitealias{Feng+2017}{F17b}
\defcitealias{Lovis+2006}{L06}
\defcitealias{Lovis+2011}{L11}
\defcitealias{Mayor+2009}{M09}
\defcitealias{Mayor+2011}{M11}
\defcitealias{PLS+11}{P11}
\defcitealias{TAE+2013}{T13a}
\defcitealias{Tuomi+2013}{T13b}

\begin{document}
  \title{New wavelength calibration of the HARPS spectrograph}

   \author{   A. Coffinet\inst{1,*},
                  C. Lovis\inst{1},
                  X. Dumusque\inst{1},
                  \and
                  F. Pepe\inst{1}
               }

   \institute{{1} Observatoire astronomique de l'Universit\'e de Gen\`eve,
              51 chemin des Maillettes, 1290 Versoix, Switzerland\\
              * \email{adrien.coffinet@unige.ch}
             }

   \date{Received XXX XX, 201X; accepted XXX XX, 201X}


  \abstract
   {Doppler spectroscopy has been used in astronomy for more than 150 years. In particular, it has permitted to detect hundreds of exoplanets over the past 20 years, and the goal today, to detect Earth-like planets, requires a precision around \SI{0.1}{\m\per\s} or better. Doppler spectroscopy has been and will be also of major importance for other studies such as the variability of fundamental constants and cosmological studies. For all these applications, having the best possible wavelength calibration is crucial. Despite the fact that the HARPS spectrograph has been operational at the 3.6-m ESO telescope for more than 15 years, and that it provides among the most precise Doppler measurements, improvements are still possible. One known problem, for instance, is the non-fully regular block-stitching of the CCDs, which introduces, in some cases, one-year period parasitic signals in the measured radial velocity.}
   {The aim of the presented work is to improve the wavelength calibration of the HARPS spectrograph to push further its planet-detection capabilities.}
   {The properties of the CCD stitching-induced pixel-size anomalies are determined with LED flat-field frames, and then a physical, gap-corrected map of the CCDs is used for the fitting model of the spectral orders. We also use a new thorium line list, based on much higher-accuracy measurements than the one used up to now. We derive new wavelength solutions for the 15 years of HARPS data, both before and after the fibre upgrade that took place in 2015.}
   {We demonstrate that we indeed correct the gap anomalies by computing the wavelength solutions of laser frequency comb exposures, known to have a very low dispersion, both with and without taking the gap correction into account. By comparing the rms of the most stable stars of the HARPS sample, we show that we globally decrease the radial velocity dispersion of the data, especially for the data acquired after the change of fibres of 2015. Finally, the comparative analysis of several individual systems shows that we manage to attenuate the periodogram power at one year in most cases. The analysis of the RVs derived from individual stellar lines also shows that we indeed correct the stitching-induced RV variation.}
   {This improved calibration of the HARPS spectrograph allows to go deeper in the search for low-amplitude radial-velocity signals. This new calibration process will be further improved by combining the thorium calibration spectra with laser frequency comb and Fabry-Perot calibration spectra, and this will not only be used for HARPS but notably also for HARPS-N and the new ESPRESSO spectrograph.}


   \maketitle
%


\section{Introduction}

\subsection{Context}
     \label{subsec:Context}

Since Isaac Newton's detailed study in 1666--1672 of the decomposition of light when it passes through a prism \citep{Newton1730}, spectroscopy has been widely used for various types of studies in physics, chemistry, biology and of course astronomy. Concerning astronomy, Doppler spectroscopy, also known as the radial-velocity technique, has been used at least since 1867, when William Huggins used this technique to determine the motions of several objects with respect to the Earth \citep{Huggins1867,Huggins1868}. Since then, this technique has led to the discovery of many star companions and, as it was suggested by \citeauthor{Struve1952} in \citeyear{Struve1952}, extrasolar planets. Since the first planetary-mass objects that were identified by Doppler spectroscopy, more than 20~years ago \citep{MQ95}, 786 of them have been revealed by this technique as of November~30th,~2018 (\citeauthor{exoplanet.eu}, see \citealt{Schneider+2011,Schneider2015,Schneider2016}), and many others have been confirmed thanks to it. Today, the precision is already better than \SI{1}{\m\per\s} \citep{PEM14}, what permits to already detect short-period \mbox{(sub-)}Earth-mass planets (see e.g.~\citealt{AD+17}), and is expected to keep improving in the coming years (see \citealt{LF10} for further historical details and \citealt{Fischer+2016} for the state of the art). The declared aim is to find Earth-like planets, which implies to be able to detect Earth-mass planets in the habitable zone of their stars. To reach this goal, the precision that is required is of the order of a few tens of centimetres per second or less (\SI{9}{\cm\per\s} for the Earth around the Sun) over timescales comparable to or longer than the orbital period.

Another recent use of Doppler spectroscopy in astronomy aims at measuring several fundamental constants that rule our universe, among which the fine-structure constant and the proton-to-electron mass ratio, and determining whether they are really constant in space and time (see e.g. \citealt{Murphy+2003,MC+13}). Also, future spectrographs are expected to be able to provide a direct and model-independent determination of the expansion history of the Universe by measuring the redshift drifts of cosmologically distant sources (the Sandage-Loeb test), which requires a precision of only few \si{\cm\per\s} over many years \citep{Sandage1962,Loeb1998,Liske+2008}. High-precision Doppler spectroscopy is also important for other astronomical studies such as, for example, stellar oscillations \citep{Arentoft+2008,Pepe+2014}.

   The \textit{High-Accuracy Radial-velocity Planet Searcher} (HARPS), operational since~2003, is a high-resolution ($R=\num{115000}$) fibre-fed cross-dispersed echelle spectrometer installed on the ESO 3.6-m telescope at La Silla Observatory, in Chile \citep{Pepe+2000,Mayor+2003}. It covers the visible spectrum from 380 to \SI{690}{\nm}. As its name indicates, it is especially dedicated to the discovery and follow-up of extrasolar planets. As of 2018, it remains one of the most precise and efficient spectrographs available in the world for radial-velocity exoplanet detection, having demonstrated a long-term precision of at least \SI{0.8}{\m\per\s} \citep{Dumusque+2012,PEM14} and having permitted to detect radial-velocity signals with semi-amplitudes down to or even below \SI{0.5}{\m\per\s} \citep{PLS+11,Dumusque+2012,FTJ17,Feng+2017}\footnote{Here we use voluntarily the word ``signals'' instead of ``planets'' given that the planethood of some of these signals is uncertain or questioned, but these signals were nonetheless detected, no matter their origin.}. Thorium has been used for the wavelength calibration of HARPS since the beginning, and improvements were added later. First came the Fabry-Perot interferometre, that became operational in 2011. The procedure for it, one step further than the present paper, is described in \cite{CC+2018}. A laser-frequency comb (LFC) was added later, that had a long period of commissioning and has just been offered to the community on November~15$^\textrm{th}$,~2018 (see \url{http://www.eso.org/sci/facilities/lasilla/instruments/harps/news.html}). Thanks to the last upgrades of HARPS, such as the use of a laser frequency comb (LFC) and octagonal fibres, the precision of the spectrograph is expected to become even better than \SI{0.5}{\m\per\s} on bright stars \citep{LoCurto+2015}. However,  the improvement of the instrumental performances has been a continuous process and some initial limitations still need to be overcome. Moreover, the goal of the present work is to improve all HARPS data, including the ``old'' ones when only the thorium lamps were available, and not only the recently acquired or future ones. These ``old'' data represent many years of data and should therefore not be put aside.

\subsection{HARPS CCD block stitching}
     \label{subsec:Intro HARPS's CCD block stitching}

\begin{figure}
    \centering
   \includegraphics[scale=0.05]{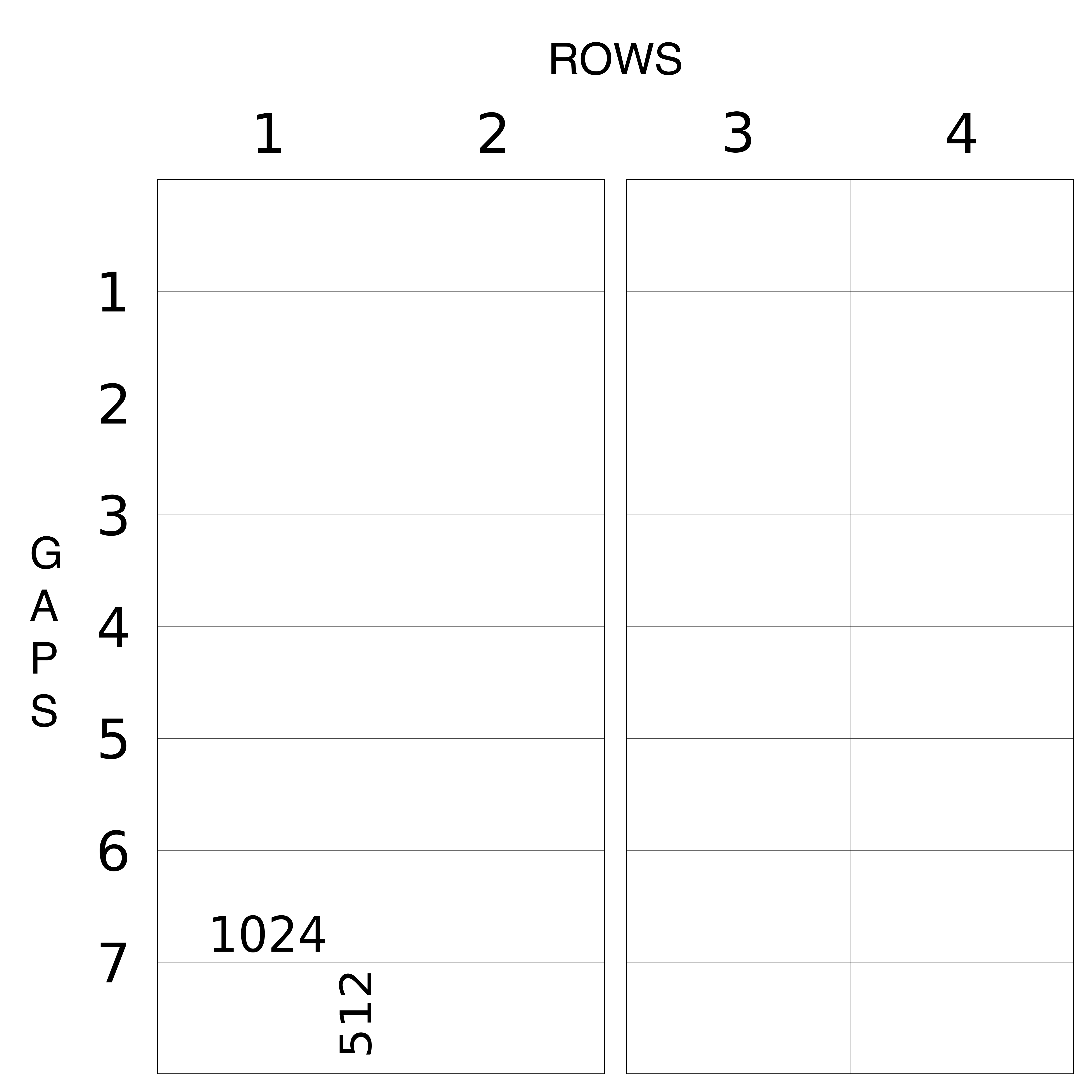}
   \caption{Schematic representation of HARPS CCDs, highlighting the stitching pattern. The row and stitching (gap) numbers are those used in Table 1. $512 \times 1024$ is the size of the blocks in number of (physical) pixels.}
   \label{fig:CCD}
\end{figure}

   The HARPS CCD (full detailled description in \citealt{ESO2001}) is a 4K$\times$4K CCD mosaic made of two 2K$\times$4K E2V CCDs type~\mbox{44-82} with a nominal pixel size of \SI{15}{\um}. As shown on Figure~\ref{fig:CCD}, these individual 2K$\times$4K CCDs are themselves made of two rows of eight $1024\,\times\,512~\mathrm{pixels}$ subparts hereafter called ``blocks'' that arise directly from the manufacturing process of the CCD. The stitching between these blocks is not perfect, such that the pixels at the edge of each of these subparts may be slightly larger or smaller than the ``central'' pixels. These inter-block pixel-size anomalies are hereafter called ``gaps''. The central pixels can also have small variations of size with respect to the nominal value, but these variations are expected to be much smaller than the inter-block gaps. Additionally, they are supposed to be random with zero average, and of such a low level that, with our technique, they cannot be distinguished from the pixel-to-pixel sensitivity variation\footnote{This could be achieved by, among other techniques, scanning the CCD with a laser frequency comb: see for example \cite{Wilken+2010}. This has been done for example by \cite{Murphy+2012} and \cite{Ravi+2017}.}, hence the fact that they will not be accounted for in our work.

The average dispersion of the HARPS spectrograph is \SI{820}{\m\per\s\per\px} (\num{2.7e-6} in relative) at the middle of the orders, i.e. \SI{0.015}{\angstrom\per\px} for a wavelength of \SI{5500}{\angstrom}. Earth's yearly revolution around the Solar System barycentre (SSB) has a peak-to-peak velocity amplitude of \SI{59.58}{\km\per\s}, i.e.~\SI{2.0e-4}{times} the speed of light in vacuum. This is equivalent to a wavelength shift of \SI{1.1}{\angstrom} at $\lambda = \SI{5500}{\angstrom}$, i.e. to a position shift of \SI{73}{pixels} on HARPS's CCD, or about one seventh of the width of a block. The 72~HARPS spectral orders are along the \mbox{512-px}-side direction, so each of them crosses 7~gaps. This makes a non-negligible part of the spectral lines imaged on the CCD come and go across the stitching boundaries with a one-year periodicity, that induces a one-year signal in the final data of a number of stars observed with this instrument. This effect has been known for a long time (see e.g. \citealt{LP07}, hereafter referred to as \citetalias{LP07}, and \citealt{DumusqueThesis}) and was already studied by several other authors (see \citealt{Wilken+2010,Molaro+2013,Dumusque+2015,BZR15}).

In order to obtain an improved solar atlas, \cite{Molaro+2013} computed new wavelength solutions of HARPS spectra. They took the block-stitching effect into account by fitting separately each 512-pixel-large section of a laser frequency comb (LFC) spectrum by a third-order polynomial. They show that, by doing this, they significantly reduce the dispersion of the residuals compared to using a single third-order polynomial for the whole order assuming that all the pixels have the same size. However, one should notice that this option implies the fit of 24~coefficients for each order, that have a priori no reason to be considered as independent. This technique is only applicable for calibration sources that produce a dense enough number of lines across all CCD blocks.

\cite{Dumusque+2015} analysed several stars and showed several examples for which the one-year signal is perfectly correlated with the barycentric Earth radial velocity (BERV), reinforcing the presumption that this one-year signal is related to the yearly motion of the Earth around the Sun. Moreover, a line-by-line analysis shows that, if the lines that cross the block stitching boundaries are removed from the analysis, the one-year signal disappears for some of the stars at least, while this one-year signal is further enhanced if one keeps only the gap-crossing lines. Despite the fact that the solution proposed by \cite{Dumusque+2015} does not solve the problem for some of the stars, indicating that the block stitching may not be the only effect (some remnant telluric lines are for example suspected; see \citealt{Cunha+2014}), it clearly demonstrates that the block stitching is at least part of the cause of the one-year signal.

One may argue that removing the gap-crossing lines is enough to solve the CCD-gaps part of the problem. However, it must be remarked that this technique will not solve the wavelength inaccuracy at its roots. Indeed, the position of the lines used for the wavelength solutions still assumes that all the pixels have the same size, hence keeping the slight inaccuracy in the fitted positions of these lines. Moreover, removing these lines reduces the number of usable stellar lines, thus reducing the precision that one could expect by using all the available lines, and given that these ``contaminated'' lines represent about 20\% of all the lines \citep{Dumusque+2015}, removing them has a clearly non-negligible effect on the reachable precision. The idea here is therefore to correct the thorium-line positions, in order to get a correct wavelength solution, and therefore radial velocities with an optimal precision, such that no stellar absorption line suffers from periodic systematic error and must therefore be excluded from the RV computation.

\subsection{New thorium spectral line dataset}
     \label{subsec:Intro dataset}

The radial-velocity technique relies on the measurement of the wavelength shift of spectral lines due to the Doppler effect and so, given that today the observed spectra are registered on CCDs, it consists fundamentally in computing the relation between a position on the CCD and the corresponding wavelength, what is called the wavelength solution. The accuracy of the latter is crucial for precise Doppler measurements. For HARPS, \citetalias{LP07} observed a dispersion of the residuals around the wavelength solutions reaching 50--\SI{70}{\m\per\s}. The origin of this high scatter is unexplained, and the knowledge of where it comes from could allow to properly correct it or at least to determine the elements that could be improved in new instruments. One can notably think about the new \SI{10}{\cm\per\s}-precision ESPRESSO spectrograph \citep{Pepe+2014,GH+17}, expected to be able to detect Earth-like planets, or further planned instruments such as the HIRES instrument expected for ESO's ELT \citep{Maiolino+2013,Zerbi+2014}.

   \cite{PE83}, hereafter \citetalias{PE83}, published the Los Alamos \textit{Atlas of the Thorium Spectrum}, a very-high-resolution ($R=\num{600000}$) catalogue of about \num{11500} thorium spectral lines between 3000 and \SI{11000}{\AA}, that had been obtained with the McMath-Pierce \SI{1}{\m} Fourier Transform Spectrometer (FTS) of the National Solar Observatory at Kitt Peak. This catalogue was used for a long time as the reference for thorium-line wavelengths and then for thorium-calibrated spectrographs (see e.g. \citet{Baranne+1996} for ELODIE or \citetalias{LP07} for HARPS). However, even at this high resolution, the provided wavelengths are affected by systematics that lie significantly above the $10^{-10}$ relative precision aimed by sub-\si{\m\per\s} RV spectrographs. Taking advantage of HARPS's stability, \citetalias{LP07} attempted to make a new list of thorium and argon spectral lines based on HARPS ThAr spectra.

\cite{RNS14}, hereafter referred to as \citetalias{RNS14}, made a new atlas that contains \num{19874}~lines of neutral, singly ionized and doubly ionized thorium (\ion{Th}{I}, \ion{Th}{II} and \ion{Th}{III}, respectively) from 250 to \SI{5500}{\nm} (\num{40 000} to \SI{1800}{\per\cm}). They experimentally measured more than 1600~thorium lines from the near ultraviolet to the near infrared with a resolution up to $10^7$ with the 2 m FTS at the National Institute of Standards and Technology (NIST), and combined them with lines observed in seven other experimental studies: \cite{GBCZ74}, \cite{ZC74}, \cite{Z76}, \cite{PE83}, \cite{Engleman2003}, \cite{LP07}, and \cite{KNS08}. \citetalias{RNS14} found that \citetalias{LP07}'s thorium line list contained a large scatter with respect to their own new list, probably caused at least partially by the HARPS CCD block stitching and the non-resolution of certain blended lines.

In this work, we use therefore the new thorium line list determined by \citetalias{RNS14} as our wavelength reference. \citetalias{RNS14}'s list contains two sets of wavelengths: the Ritz wavelengths, that were computed from globally optimized energy levels, and the measured wavelengths. According to \citetalias{RNS14}, the Ritz wavelengths have a much lower uncertainty than the measured wavelengths. This is due to the fact that the Ritz wavelengths take the observed ones as input to an energy level model of the atom/ion in question, so the individual wavelengths are then effectively constrained by many observed lines. We therefore used these Ritz wavelengths as our reference.

\subsection{The 2015 HARPS upgrade}
    \label{subsec:Intro HARPS's 2015 change of fibres}

In May~2015, the circular fibres of HARPS were replaced by octagonal fibres \citep{LoCurto+2015}. This change had a noticeable impact on the line profile (e.g. FWHM) and a spectral-type dependent RV offset could be observed between before and after the change of fibres. Up to now, the wavelengths used for the post-2015 data were those that had been computed by \citetalias{LP07} for the pre-2015 HARPS. These were therefore not optimised for this period and the wavelength solutions were consequently not as good as for the pre-2015 data. This could be seen as degraded residuals around wavelength solutions and an increase in the rms of the computed radial velocities. For these reasons, HARPS before and after the change of fibres was considered as two different instruments. The wavelength calibration improvement presented in this paper was therefore done twice independently, and the data analysis also considers the pre-2015 and post-2015 data as two separate datasets.

\section{Calibration data sets}
     \label{sec:Calibration data sets}

Our work uses and is based on different calibration data sets, each one for a specific task. Hereafter we describe them for a better overview.


\subsection{LED flat-field frames}
     \label{subsec:LED frames}

\begin{figure}
    \centering
   \includegraphics[scale=0.4]{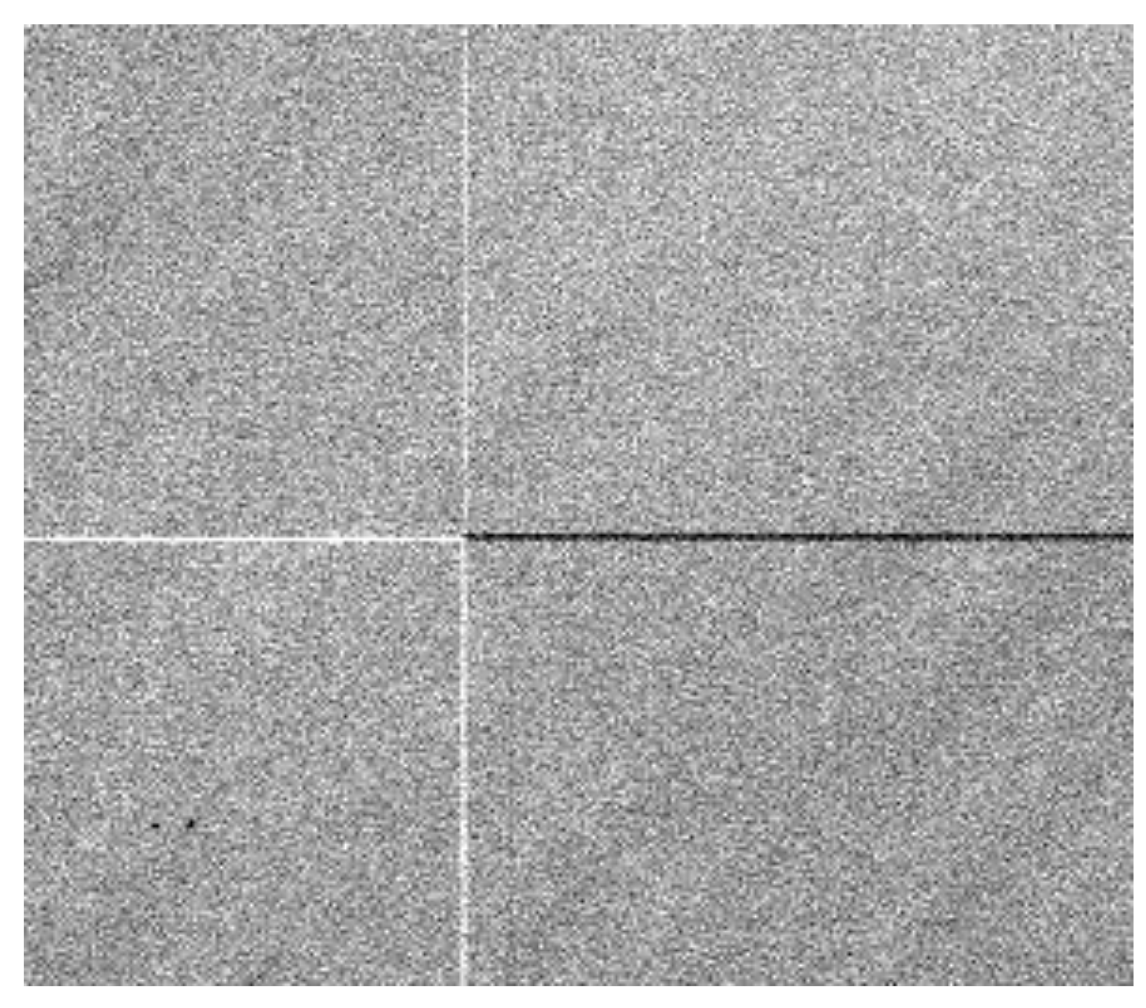}
   \caption{LED flat-field image of a part of a CCD highlighting gaps at block boundaries.}
   \label{fig:Image_Gap}
\end{figure}

The first type of frames that was used is LED flat-field frames. These frames record the response of each individual pixel under a homogeneous illumination, and are then used to correct pixel-to-pixel variation effects.

Ten such frames were used for the determination of the positions and the sizes of the gaps (see Sections~\ref{subsec:Positions of the gaps} and \ref{subsec:Sizes of the gaps} respectively). The interest of using LED flat-field frames here is that they provide an information that is independent from the wavelength calibration frames (that use a spectral source), so a consistency check can be made after correction.

As presented in Figure~\ref{fig:Image_Gap}, on these LED frames, we unsurprisingly see pixel-to-pixel brightness variations. These brightness variations are usually considered to be due to pixel-to-pixel sensitivity variations (technically referred to as Photo-Response Non-Uniformity, PRNU), assuming that all the pixels have the same size, and flat-field frames are actually usually used to correct these sensitivity variations. However, these brightness variations can also be due to effective differences of the collecting area of the pixels, for instance at the boundary between two blocks, if the electrode pattern produced by photo-lithography had not been replicated with a shift that is exactly a multiple of the pixel size. A priori, we cannot discriminate between a sensitivity difference produced by variable quantum efficiency or by a different size of the pixel. Nevertheless, a regular pattern located at the boundaries between the blocks would be a clear indication for the latter: in fact, we find every 512~pixels a column with a brightness significantly different from its neighbours. This brightness difference is much larger than the ``normal'' local pixel-to-pixel brightness scatter that we can see everywhere else on the frame, and the regularity of the pattern reinforces the idea that it is not due to a white noise. These columns are actually four to eight 1024-pixel-high sub-columns (four times 1024~pixels in height, and one or two pixels in width), each sub-column being darker or sometimes brighter at different levels. These are actually known since the beginning of HARPS (see e.g. \citealt{ESO2001}, where a picture is shown in section~1.1.16, page~13), and have always been strongly suspected to be due (at least partially) to the gaps that we want to characterise.


\subsection{Laser-frequency-comb calibration frames}
     \label{subsec:Laser frames}

A set of 82~laser-frequency-comb (LFC) calibration frames taken with HARPS between 2015-04-17T02:57:32.992 and 2015-04-17T04:20:07.352, i.e. not long before the upgrade of May~2015 \citep{LoCurto+2015}, was used to cross-check the sizes of the gaps estimated from the LED exposures. The advantage of laser frames compared to atomic-lamp frames (in our case, thorium-argon frames, see~Section~\ref{subsec:ThAr frames}) is that the absolute wavelength of each line of the laser frequency comb is known with an extreme precision. Indeed, only two parameters are needed to describe the wavelengths of the whole set of laser lines: $\nu_0$, the anchor frequency of our laser comb, and $\Delta\nu$, the mode spacing. Then, the formula that gives the wavelength of a given line is $\nu_n = \nu_0 + n \cdot \Delta\nu$, for every integer $n$. In our case, $\nu_0 = \SI{288.059800}{\THz}$ and $\Delta\nu = \SI{18.0}{\GHz}$ (Lo~Curto, private communication). Knowing the absolute wavelength of a single laser line for each order therefore gives the wavelengths of all the laser lines recorded on the CCD. 

Unfortunately, the laser comb does not currently cover the whole CCD; only the orders between 26 and the last one (order~72) could then be used. This covers about three quarters of the whole CCD, especially three of the four 1024-pixel-high rows of CCD blocks, the fourth one being only covered by a single order (order~26).


\subsection{Thorium-argon calibration frames}
     \label{subsec:ThAr frames}

Three sets of thorium-argon calibration frames were used during this process. Those lamp frames are used as the absolute wavelength calibrators of our data. The first two sets correspond to reference calibration data of HARPS before and after its upgrade in May~2015 \citep{LoCurto+2015}, respectively. The first set is composed of 64~frames of December~2004, all with thorium-argon on both fibres, and therefore usable for the calibration of both fibres. The second one consists of 120~frames of June~2015, distributed as such: 48 frames with thorium-argon of both fibres, usable for the calibration of both fibres, and 72 with thorium-argon on fibre~A and Fabry-Perot on fibre~B, therefore usable only for the calibration of fibre~A. Finally, the third set contains 10~frames from April~13th to April~23rd,~2015, used to compare the results with the ones obtained with the laser frames, that are of April~17th,~2015 (see~Section~\ref{subsec:Laser frames}).

The spectral lines of the first two sets were fitted and matched to our list of thorium lines, whose approximate positions on the detector are known. We got lists of lines for both fibres, both before and after the upgrade of May~2015. Before the upgrade, we counted for fibre~A a list of \num{10227}~lines, and for fibre~B a list of \num{10006}~lines, most of them being either thorium lines or argon lines. After the upgrade, the lists were reduced to \num{9944}~lines for fibre~A and \num{9348}~lines for fibre~B. The difference is due to the fact that some lines that were well fitted for before the fibre upgrade are no more well fitted after (mainly blended lines).

\section{Determination of the properties of the gaps}
     \label{sec:gaps}


\subsection{Positions of the gaps}
     \label{subsec:Positions of the gaps}

To determine the positions of the gaps, we used the LED flat-field frames presented in Section~\ref{subsec:LED frames}. A spatially regular brightness anomaly, large in intensity compared to the ``normal'' pixel-to-pixel brightness variation present on the rest of the image, is easily visible by eye, as shown on Figure~\ref{fig:Image_Gap}. We notice that, as already stated by \citet{Molaro+2013}, the positions of the large brightness anomalies are offset from exactly the multiples of 512. \citet{Molaro+2013} noted an offset of 4~pixels between the physical positions of the gaps on the CCD (at $n_j~=~512~j~\mathrm{pixels}$) and the positions observed on the flat fields (at $n_j~=~(512~j+4)~\mathrm{pixels}$), and that 1 or 2~pixels are involved at each time. Here we find 3 or/and 4~pixels of offset, given that several of the anomalies are on two pixels of width. Indeed, the pixels that we see to be affected are 515-516, 1027-1028, 1539-1540, 2051-2052, 2563-2564, 3075-3076, and 3587-3588 (when numbering the pixels of the CCD from 1 to 4096).


\subsection{Sizes of the gaps}
     \label{subsec:Sizes of the gaps}

Obviously, even if the abovementioned flat-field frames already have all individually high signal-to-noise ratios (SNR), the more frames can be accumulated, the more precise the measurement of the gap will be. We therefore stacked the ten LED flat-field frames mentioned in Section~\ref{subsec:LED frames} into one ``master'' flat-field frame. 

The first hypothesis that was made is that the gaps have a constant size between two given ($512\times1024$)-pixel blocks. This may not be strictly true; however, the variation of the size of a gap along the border between two given blocks is expected to be very small. This assumption corresponds to assuming that the tilt between two neighbouring blocks is negligible. Practically, no significant variation in the brightness anomaly is observed along the border between two blocks, that makes us confident in the fact that this hypothesis is satisfied here.

Secondly, we assume that, by default, the pixels have a nominal size $L_\mathrm{nom}$, equal to \SI{15}{\um}, in both directions. The only size anomaly that we take into account is the one that we want to characterise, i.e. the gap size anomaly along the spectral orders. This assumption directly comes from the already-mentioned fact that one cannot distinguish the pixel-to-pixel intrinsic sensitivity variation from the relative pixel size anomaly, and by the fact that, as we can see on Figure~\ref{fig:Image_Gap}, the amplitude of the gap-induced CCD brightness anomaly is much larger than the ``normal'' pixel-to-pixel brightness variation. The ``size'' of the pixels will therefore now refer only to their size along the direction of the spectral dispersion, and the brightness anomalies of the columns affected by the gaps are therefore considered to purely reflect their pixel-size difference along the direction of the orders, and therefore the size of the gaps.

Following the first hypothesis, we cumulated together blocks of 1024~lines of height as one single line, what increases significantly the SNR. One should notice that, assuming a random distribution of the pixel characteristics (size, sensitivity, ...), this line stacking allows to significantly reduce the variation in size in the direction of the cross-dispersion, reinforcing the justification of the second hypothesis. For the same reason, this stacking also significantly reduces the ``normal'' pixel-to-pixel sensitivity variation in the spectral direction. At the end, we therefore got a ``master flat-field frame'' made of four 4096-pixel-long lines, each line containing 7 one-to-two-pixel-large ``large'' brightness anomalies.

The size of the gaps, relative to the nominal size of a pixel, was therefore calculated as:

\begin{eqnarray}
\frac{L_\textrm{gap}}{L_\textrm{nom}} = \frac{L_\textrm{est}}{L_\textrm{nom}}-1 = \frac{F_\textrm{meas}}{F_\textrm{exp}}-1
     \label{eqn:gapsize}
\end{eqnarray}

where $L_\textrm{est}$ is the estimated actual size of the affected pixel, $F_\textrm{meas}$ is the flux actually measured on this pixel, and $F_\textrm{exp}$ is the value that we would expect if the pixel had a nominal size.

One should notice that, with this convention, the negative $L_\textrm{gap}$ correspond to pixels smaller than nominal, i.e. to a pixel darker than ``normal'', while the positive $L_\textrm{gap}$ correspond to pixels larger than nominal, i.e. to pixels brighter than ``normal''.

The expected flux was computed by interpolating the flux of neighbouring pixels with a second-order polynomial. In each side of a given pixel $n$ for which the gap size was to be computed, the pixels selected for the interpolation were $[n-10,n-2[\cup]n+2,n+10]$.

We therefore got the estimated sizes of all the gaps for the four rows. The values that we obtained are in Table \ref{tab:gapsize}. For each gap, there are two values corresponding to the correction of the two pixels that are each time affected. The maximum values for individual pixels are \SI{-34.54}{mpx} and +\SI{15.27}{mpx}, corresponding to \SI{-28.3}{\m\per\s} and +\SI{12.5}{\m\per\s}. Over two pixels corresponding to the same gap, the maximum values are \SI{-62.00}{mpx} and +\SI{25.76}{mpx}, i.e. \SI{-50.8}{\m\per\s} and +\SI{21.1}{\m\per\s}.

We can compare these values with those obtained by \cite{BZR15}. All except 4 are within 6~mpx ($\approx \SI{5}{\m\per\s}$) of the values found by \cite{BZR15}; most of them are within 3~mpx ($\approx \SI{2.5}{\m\per\s}$). The average and the dispersion of the differences (\citealt{BZR15} minus our values) are respectively -0.17~mpx ($\approx \SI{-0.14}{\m\per\s}$) and 6.23~mpx ($\approx \SI{5}{\m\per\s}$).

\begin{table}[h]
\centering
\begin{tabular}{|c|r|r|r|r|}
  \hline
  block & \multicolumn{4}{c|}{Estimated gap sizes [$\times10^{-3} L_\mathrm{nom}$]} \\
  \cline{2-5}
  group & \multicolumn{1}{c|}{1} & \multicolumn{1}{c|}{2} & \multicolumn{1}{c|}{3} & \multicolumn{1}{c|}{4} \\
  \hline
  1 &   -9.92 & -19.67 & -29.72 & -26.34 \\
     & -18.28 & -30.71 &  +3.33 & -16.12 \\
  \hline
  2 &   -5.69 & +13.48 &  +5.20 &  +8.97 \\
     &  +1.39 & +12.28 & -14.36 &  +2.96 \\
  \hline
  3 & -18.30 &   -9.14 & -11.02 &   -1.57 \\
     & -19.62 & -16.73 &   -4.16 &  - 6.99 \\
  \hline
  4 &  +0.88 & -12.49 &  +5.02 & -16.08 \\
     &  +0.28 & -17.63 & -15.62 & -21.25 \\
  \hline
  5 &  +0.26 &   -1.92 &   -9.77 &  +3.45 \\
     &   -8.46 &   -5.86 &   -4.52 &  +1.90 \\
  \hline
  6 & -23.84 & -17.08 & -27.46 & -13.80 \\
     & -29.01 & -24.26 & -34.54 & -32.18 \\
  \hline
  7 & -15.79 &  +6.88 &  +5.04 & +15.27 \\
     &   -5.77 &   -6.46 &   -8.16 &  +6.38 \\
  \hline
\end{tabular}
\caption{Estimated gap sizes. $L_\mathrm{nom} = \SI{15}{\um}$.}
     \label{tab:gapsize}
\end{table}

Since we know where the spectral orders are on the CCD, we can therefore make a ``pixel-size map'' of the order-by-order, pixel-by-pixel correction to apply to the (extracted) spectra. This map has therefore an e2ds [extracted 2d spectrum] format, i.e. 72~lines of 4096~values. This map will be used later when we will compute the new wavelength solutions. Here is the process we used to build this map:

\begin{itemize}
  \setlength\itemsep{1em}
  \item For the orders that belong to only one of the four rows of CCD blocks (what is actually the case for all but one order), the correction is easy to find since it is simply the correction of the corresponding row.
  \item However, for the orders that cross the border between two rows, we use an averaged gap size weighted by the number of pixels extracted in each row at the given position on the order.
\end{itemize}

\section{New wavelength solutions}
     \label{sec:New wavelength solutions}

In an analogous way as \citetalias{LP07}, the idea was to fit the wavelengths of the spectral lines as a polynomial of their positions on the CCD. The difference between \citetalias{LP07} and the present work is that here the gap sizes are taken into account in the equations and that we use the new thorium wavelengths of \citealias{RNS14}.

\subsection{Mathematical formula}
     \label{subsec:Mathematical formula}

 The basic idea is to keep \citetalias{LP07}'s general formula. For a given order $o$, and a given line $i$ of this order, the equation is:
\begin{eqnarray}
\sum\limits_{k=0}^P a_k X_i^k = \lambda_i
     \label{eqn:formulelp07}
\end{eqnarray}

where:
\[
      \begin{array}{lp{0.8\linewidth}}
         \{a_k\}_{k \in [0,P]}                & \textit{unknown} coefficients of the polynomial of the wavelength solution of order $o$. $P$ is the fixed degree of the polynomials, identical for all the orders    \\
         X_i                 & true position (in whatever physical unit, it be microns or nominal pixels for example) of line~$i$ on the CCD along the order ($x$-axis) \\
         \lambda_i       & (known) wavelength of line $i$ from \citetalias{RNS14}
      \end{array}
   \]
   
   The $X_i$ of this formula is supposed to be the true position of the line on the CCD, actually corresponding to the nominal-pixel-scaled pixel-number position plus the gaps:
\begin{eqnarray}
X_i = x_i + \sum\limits_{j=1}^G \delta_{i,j} g_j
     \label{eqn:conv}
\end{eqnarray}

where:
\[
      \begin{array}{lp{0.8\linewidth}}
         x_i                 & (known) ``false'' position of line $i$ on the CCD along the order ($x$-axis) as measured in terms of pixels. Practically, it simply means that this position does not take the gaps into account.\\
         G                   & (known) number of gaps on line $i$'s order \\
         \delta_{i,j}      & (known) coefficients with value 0 or 1 depending on the relative position of line $i$ with respect to gap number $j$: noting $n_i$ and $n_j$ the respective (pixel) positions of line~$i$ and gap~$j$, $\delta_{i,j}~=~\left\{
             \begin{array}{ll}
                 1 & \mathrm{if}~n_i > n_j \\
                 0 & \mathrm{else}
             \end{array}
         \right.$             \\
         g_j                 & size of gap number $j$
      \end{array}
   \]

The $\{n_j\}_{j \in [1,G]}$ and the $\{g_j\}_{j \in [1,G]}$ are those determined respectively in Sections~\ref{subsec:Positions of the gaps} and ~\ref{subsec:Sizes of the gaps}.

When we insert Formula~\eqref{eqn:conv} inside Formula~\eqref{eqn:formulelp07}, we get:
\begin{eqnarray}
\sum\limits_{k=0}^P a_k \left(x_i + \sum\limits_{j=1}^G \delta_{i,j} g_j \right)^k = \lambda_i 
     \label{eqn:equationnew}
\end{eqnarray}

The only unknowns are the $\{a_k\}_{k \in [0,P]}$, and equation~\eqref{eqn:equationnew} is then linear with respect to the unknowns. For simplicity, we will from now on use the $X_i$ (= $x_i + \sum\limits_{j=1}^G \delta_{i,j} g_j)$ in the writing of the equations.

Practically, equation~\eqref{eqn:equationnew} will be different depending on the type of line that we are dealing with. We have 4 types of lines: the thorium lines that appear on two different orders (``duplicate thorium lines''), the thorium lines that appear on only one order (``unique thorium lines''), the non-thorium lines that appear on two orders (``duplicate non-thorium lines'') and the non-thorium lines that appear on only one order (``unique non-thorium lines'').

\begin{enumerate}  
\item Duplicate thorium lines:
	\begin{eqnarray}
	\sum_k a_k X_i^k - \Delta\lambda_i = \lambda_{i,\;{\rm RNS14}}
		\label{eqn:eqTh2}
	\end{eqnarray}

\item Unique thorium lines:
	\begin{eqnarray}
	\sum_k a_k X_i^k = \lambda_{i,\;{\rm RNS14}}
		\label{eqn:eqTh1}
	\end{eqnarray}

\item Duplicate non-thorium lines:
	\begin{eqnarray}
	\sum_k a_k X_i^k - \lambda_i = 0
		\label{eqn:eqNT2}
	\end{eqnarray}

\item Unique non-thorium lines: no equation

\end{enumerate}

In these equations, the $\Delta\lambda_i$ is the unknown systematic offset of line~$i$ (for thorium lines),  $\lambda_{i,\; {\rm RNS14}}$ is the \citetalias{RNS14} wavelength of line~$i$ (for thorium lines) and $\lambda_i$ is the unknown wavelength of line~$i$ (for unidentified or argon lines). Argon lines are included in the group of lines with unknown wavelengths because they do not appear in \citetalias{RNS14}'s list and because they are significantly more sensitive to physical conditions in the lamp than the thorium lines (see \citetalias{LP07}). As in \citealias{LP07}, the $\Delta\lambda_i$ is the same for both occurrences of a given duplicate thorium line, since systematic effects should affect in the same way these two occurrences. Coupling the orders in this ways also strengthen the reliability of the wavelength solutions towards the order edges, where the wavelength solutions are usually more poorly constrained.

\subsection{Thorium-line systematics}
     \label{subsec:Thorium-line systematics}

\begin{figure}
    \centering
   \includegraphics[scale=0.5]{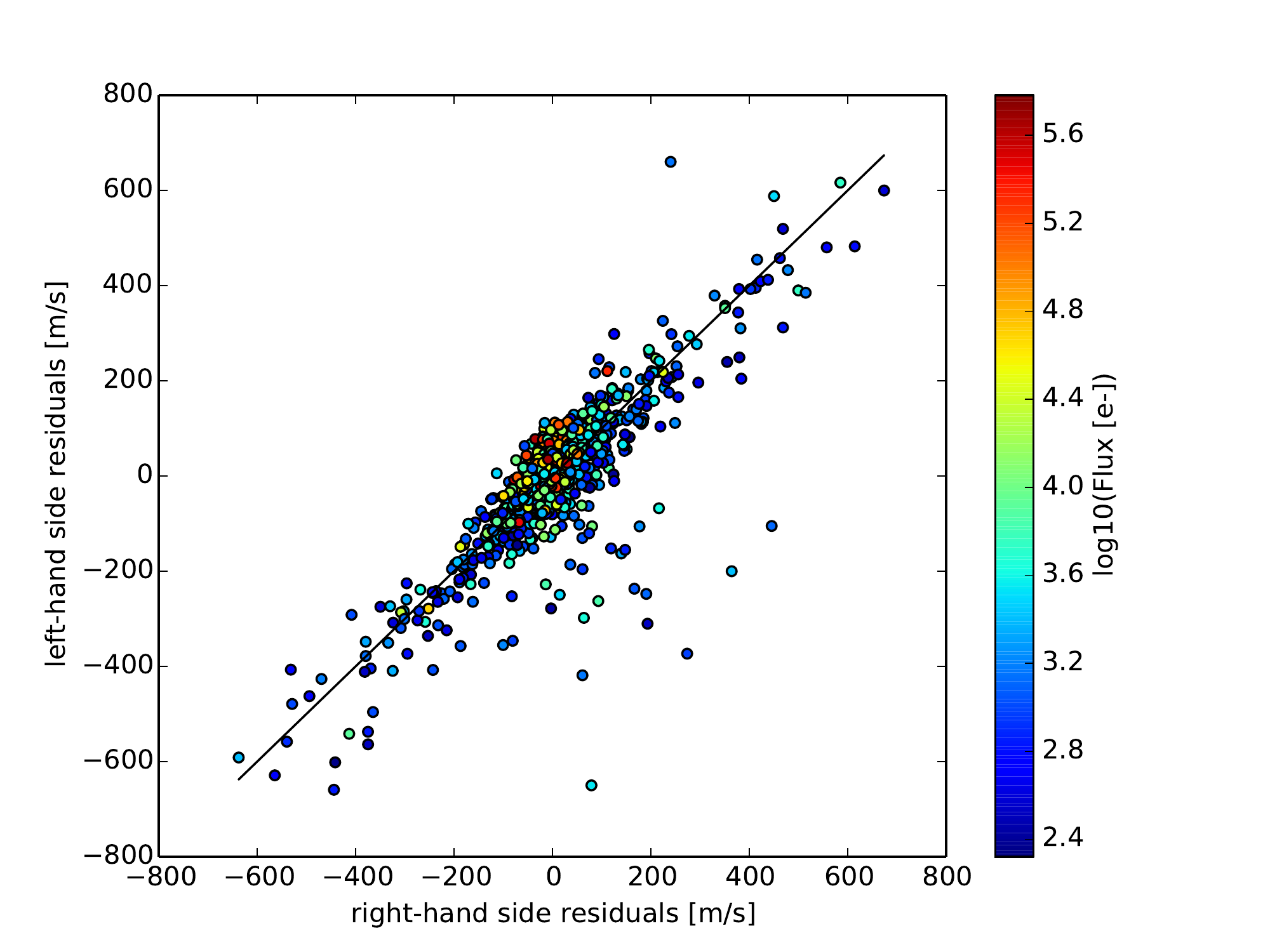}
   \caption{Residuals of the duplicate thorium lines. The black line is the line of equation $y=x$.}
   \label{fig:left-right}
\end{figure}

The individual systematics $\Delta\lambda_i$ are unknowns to be determined. To understand their origin, we computed the wavelength solutions for each individual order based on \citetalias{RNS14}'s wavelengths. Only thorium lines are therefore used, both unique and duplicate, given that only thorium lines are present in \citetalias{RNS14}'s catalogue. Since the orders are decoupled, the system contains only equations of type~\eqref{eqn:eqTh1}. The residuals of the two occurrences of the duplicate thorium lines (the left-hand-side occurrence being the one of the order following the one of the right-hand-side occurrence) are plotted against one another in Fig.~\ref{fig:left-right}, with a superimposed 1-to-1 line. We clearly see that there is a good 1-to-1 correlation between the left-hand-side residuals and the right-hand-side ones. Moreover, we can see that the distribution of the residuals is centered on zero. From this observation, two things can be deduced: first, the residuals (so basically the $\Delta\lambda_i$) are on average equal to zero; second, the $\Delta\lambda_i$ of the equations corresponding to a same duplicate line should all be considered to be equal. In addition, it means that the $\Delta\lambda_i$ represent systematics that are not related to the CCD, the optics or the PSF, but rather are intrinsic to the thorium spectrum; they could for example be related to blends that are not resolved by HARPS. These elements will add significant constraints on the $\Delta\lambda_i$. These global properties of the $\Delta\lambda_i$ translate into these additional equations (one for each $\Delta\lambda_i$, i.e. one for each duplicate thorium line):

\begin{eqnarray}
\Delta\lambda_i = 0
	\label{eqn:eqDelta}
\end{eqnarray}

This has to be understood in a least-squares sense: these equations will ensure that, in the global solution, the properly normalised (see below)  $\Delta\lambda_i$ will have a mean value close to zero and a variance close to one. The weight of these equations should be representative of the dispersion of the corresponding residuals (cf.~Section~\ref{subsec:Weighting of the equations}).

\subsection{Weighting of the equations}
     \label{subsec:Weighting of the equations}

\begin{figure}
    \centering
   \includegraphics[scale=0.5]{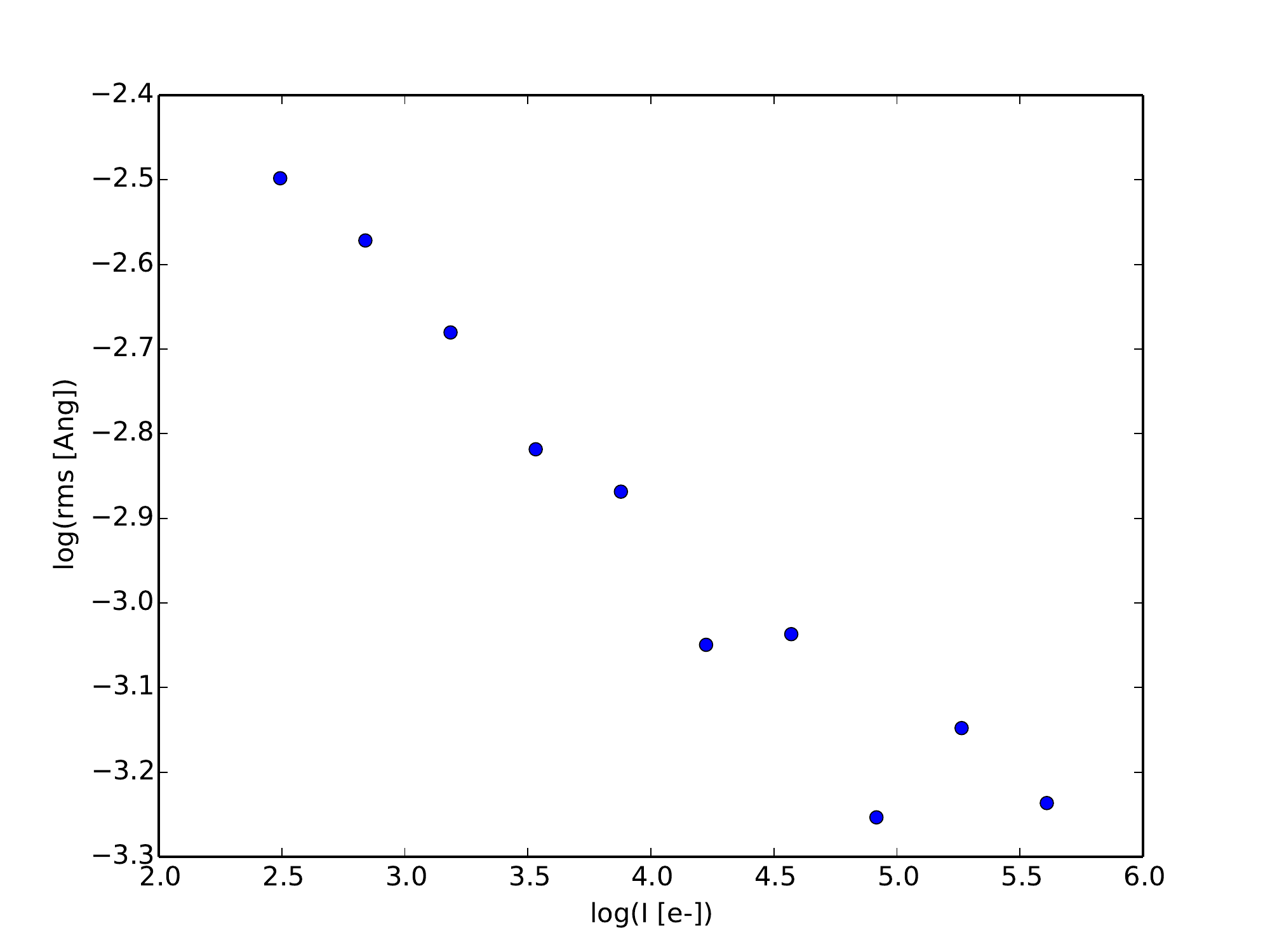}
   \caption{Dispersion of the thorium-line systematic offsets $\Delta\lambda_i$ (in \AA) vs. logarithm of the line intensity (in electrons).}
   \label{fig:log-log}
\end{figure}

We need to assign the proper weight to each equation in order to properly solve our weighted linear least-squares problem. Uncertainties for each equation will be computed differently depending on the equation type. For the equations of types~\eqref{eqn:eqTh2} and \eqref{eqn:eqNT2}, uncertainties are the quadratic sum of photon noise (considered here as the wavelength-converted uncertainty on the fitted position of the line) and a fixed detector-related noise (0.01 pixel). This detector-related noise is a noise floor such that we cannot measure the position of a line with an accuracy better than this value. This value of 0.01~px was chosen because, as we will see in Section~\ref{subsubsec:Results laser} and Figure~\ref{fig:laser_with_or_without_gap}, the residuals of the laser wavelength solutions have a rms of about this value, whose origin is most likely instrumental (e.g.~intra-pixel sensitivity variations) and not from the (much lower) photon noise. For the equations of type~\eqref{eqn:eqTh1}, uncertainties are given by the quadratic sum of the uncertainty on the systematic offset $\Delta\lambda_i$, the line-position uncertainty and \citetalias{RNS14}'s wavelength uncertainty. Finally, equations of type~\eqref{eqn:eqDelta} are weighted by the quadratic sum of the uncertainty on the systematic offset and the \citetalias{RNS14} wavelength uncertainty.

To determine the true uncertainty of each line (for single lines) or couple of lines (for actually duplicate lines), we can go back to Fig.~\ref{fig:left-right}. In addition to comparing the left-hand-side residuals to the right-hand-side ones of the duplicate thorium lines, this figure compares these residuals to the actual intensity (flux) of these lines. One has to note that, here, the flux are the flux measured by HARPS, corrected from the blaze and average over the two occurrences. One can see that the dispersion of the residuals seems to be lower for lines with higher intensities. More specifically, if we plot the rms of the residuals with respect to the flux, we find a linear relation between the logarithm of both quantities. This is shown on Fig.~\ref{fig:log-log}, where we did 10~bins of equal size on the log(flux) scale. This correlation allows us to use this flux-based fitted rms as the uncertainty on the systematic offset $\Delta\lambda_i$ mentioned before. This makes sense given that only thorium lines have equations using this uncertainty (equations of types~\eqref{eqn:eqTh1} and \eqref{eqn:eqDelta}).

We fitted a linear relation between log(rms) and log(flux) and we therefore get the following relation between the uncertainty on the systematic offset of the thorium lines and their flux:

\begin{eqnarray}
\log(\sigma~[\r{A}]) = -0.2488 \times \log(I~[\mathrm{e}^{-}]) - 1.8311
	\label{eqn:eq-rms-flux-log}
\end{eqnarray}

i.e.

\begin{eqnarray}
\sigma~[\r{A}] = 0.01475 \times (I~[\mathrm{e}^{-}])^{-0.2488}
	\label{eqn:eq-rms-flux}
\end{eqnarray}

For a line at \SI{5500}{\angstrom}, this relation corresponds to an uncertainty on the systematic offset equivalent to \SI{25}{\m\per\s} for a flux of 1~million~electrons and to about \SI{150}{\m\per\s} for a line with only 1000~electrons. This is similar to what \citetalias{LP07} had found for their flux-dispersion relationship, where they got uncertainty between 20 and \SI{160}{\m\per\s}. This shows that the possible errors in the wavelengths of \citetalias{PE83}, assumed by \citetalias{LP07} to be the cause of these systematic corrections, can only be part of it and that necessarily additional causes are also responsible of these systematics. One can think of line blends due to the lower resolution of HARPS compared to \citealias{RNS14} or other systematic effects.

\subsection{Littrow test}
     \label{subsec:Littrow test}

From the spectrograph optical design, the wavelength at a given position along the main dispersion must follow a smooth relation from an order to the next. From the grating equation, we can write that, for a given $x$, this relation is:

\begin{eqnarray}
\lambda = \frac{K}{m}
	\label{eqn:eq-littrow1}
\end{eqnarray}

where $K$ is a constant and $m$ is the physical number of the echelle order (161-89 except 115 for fibre~A; 161-89 except 116 and 115 for fibre~B). In order to include optical aberrations, we decided to actually use a third-degree polynomial:

\begin{eqnarray}
\lambda = \sum_c b_c \left( \tfrac{1}{m} \right )^c
	\label{eqn:eq-littrow2}
\end{eqnarray}

For a given $x$ and a given order, we know that this point must satisfy both the wavelength solution of its order (equation~\ref{eqn:formulelp07}) and the Littrow equation of its position (equation~\ref{eqn:eq-littrow2}):

\begin{eqnarray}
\sum_k a_k X^k - \sum_c b_c \left( \tfrac{1}{m} \right )^c = 0
	\label{eqn:eq-littrow3}
\end{eqnarray}

In order to constrain more strongly the wavelength solutions, we add to the system of equations these Littrow equations for all the orders at five different positions (the same five positions for all the orders). We gave the same weight $\frac{1}{\sigma_\mathrm{L}^2}$ to these $N$ equations. The cumulated weight of these equations was set to be equal to the total weight coming from the addition of the weights corresponding to the uncertainty on the position of all the lines $i$ used:

\begin{eqnarray}
N \times {\frac{1}{\sigma_\mathrm{L}^2}} = \sum_{i} {\frac{1}{\sigma_i^2}}
	\label{eqn:eq-littrow-sig}
\end{eqnarray}

The corresponding uncertainty attributed to these equations is therefore:

\begin{eqnarray}
\sigma_\mathrm{L} = \sqrt{ \frac{N}{\sum\limits_i (1/\sigma_i^2)} }
	\label{eqn:eq-littrow-sig}
\end{eqnarray}

We chose to give the same weight to each of these equations because they are not directly linked to the line content of the spectrum.There is therefore \textit{a priori} no reason to favor one position over another. Also, the overall weight of these equations was set such that it does not ``hide'' the other equations, but also such that it has a noticeable influence. From a ``mechanical'' point of view, there is no reason that one direction (along the ordres) be privileged with respect to the other (across the ordres). Equating the overall weight of these two types of equations therefore reflects this.

\subsection{New wavelength solutions}
     \label{subsec:Results thorium}

Before solving the linear least-squares problem, we first built a master line list from the set of thorium frames described in Section~\ref{subsec:ThAr frames}. Measured line positions are averaged over a large number of frames to minimise photon-noise errors on line positions. The system of equations was then solved a first time to identify outliers, and then a second iteration was done with only the lines that were less than $7~\sigma$ away from the solution of the first run. This is the solution of this second run that was used for the final attribution of wavelengths. For the dataset of December 2004, we get a final reduced chi-square of 1.02 for fibre~A and 0.81 for fibre~B. For the dataset of June 2015, we get a final chi-square of 1.42 for fibre~A and 1.07 for fibre~B.

The new wavelengths obtained from this procedure are then saved in a new reference line list that is used to reprocess all individual wavelength calibration exposures acquired with HARPS over 15~years.  This new list is specific to HARPS and contains the updated wavelengths that include the line systematics $\Delta\lambda_i$. The typical dispersion of the residuals around the new wavelength solutions is \SI{10}{\m\per\s} for the pre-upgrade data and \SI{15}{\m\per\s} for the post-upgrade data. While the pre-upgrade solutions show similar dispersions compared to the old reduction, we observe a significant improvement for the post-upgrade solutions (from 30 to \SI{15}{\m\per\s}).

\section{Results}
     \label{sec:Results final}

\subsection{Test of the gap correction with the laser frequency comb}
     \label{sec:Test of the gap correction with the laser frequency comb}


\subsubsection{Principle}
     \label{subsubsec:Principle laser}

We used laser-frequency-comb frames to check our gap size estimates. The advantage of laser frames is that the absolute wavelength of each line is known with an extreme precision. When we do the wavelength solution by fitting a polynomial $P$ such that $\lambda = P(x)$, where $\lambda$ is the wavelength and $x$ is the position \textit{in number of pixels}, the gaps should cause jumps in the residuals, given that ``pixel number'' is not a physical unit of length (given the pixel-size variations). On the contrary, if we take into account the gaps by using the \textit{actual physical position} of the lines as the $x$ coordinate, no jump should remain. Obviously, the gaps would be visible only if their effect is larger than the dispersion of the residuals; however, for the laser, the dispersion is expected to be very small, hopefully smaller than the effect of the gaps (the value of this dispersion is checked afterwards). This is why we used these laser frames to check the estimated values and not the thorium frames, that are known to have a much larger dispersion, about 50--\SI{70}{\m\per\s}, larger than the RV-converted size of the gaps. The set of 82~laser frames mentioned above (Section~\ref{subsec:Laser frames}) was therefore used to perform this test.

As mentioned above, the laser comb does not currently cover the whole CCD, but only the orders from 25 to 69. Three of the four 1024-pixel-high rows of CCD blocks are fully covered, while only one single order is on the fourth one. This could limit the consistency of our verification; however, it seems reasonable to assume that, if the estimated sizes of the gaps look consistent for all the orders of the three fully-covered rows, and that the single order of the fourth raw also gives consistent results, the full set of size estimates is consistent with the actual sizes of the gaps. This is therefore what we considered in the rest of this study.


\subsubsection{Results}
     \label{subsubsec:Results laser}

\begin{figure*}
\centering

\subfloat[Order~25, without gap correction.]{\includegraphics[width=0.49\textwidth, keepaspectratio]{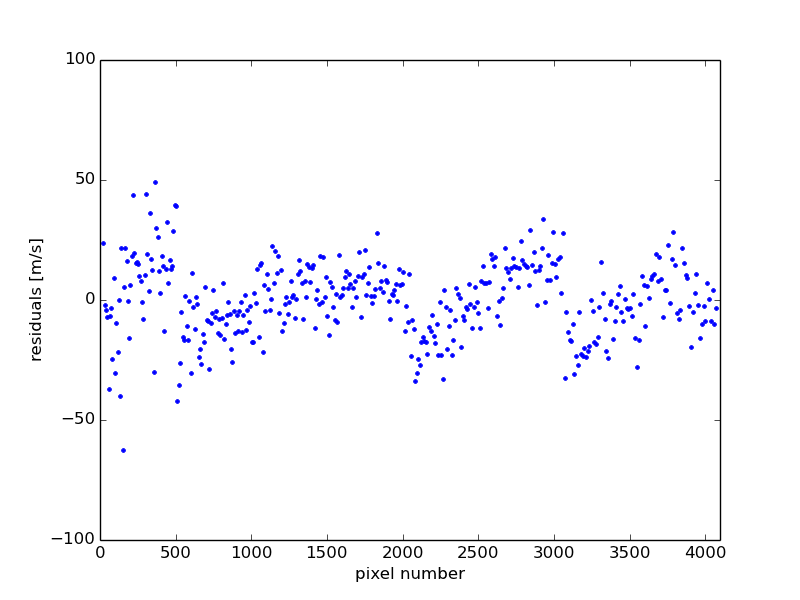}\label{fig:laser_without_gap_ordre_25}}
\hspace*{0.02\textwidth}
\subfloat[Order~25, with gap correction.]{\includegraphics[width=0.49\textwidth, keepaspectratio]{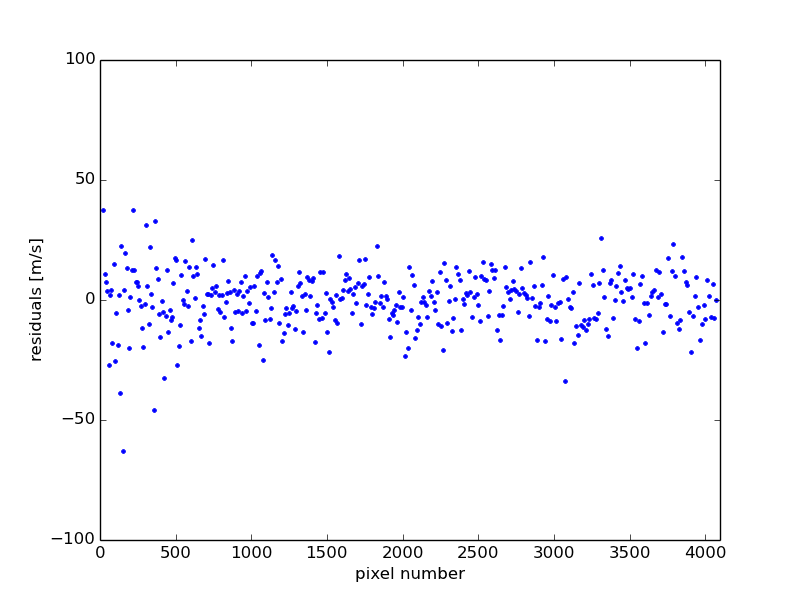}\label{fig:laser_with_gap_ordre_25}}

\subfloat[Order~42, without gap correction.]{\includegraphics[width=0.49\textwidth, keepaspectratio]{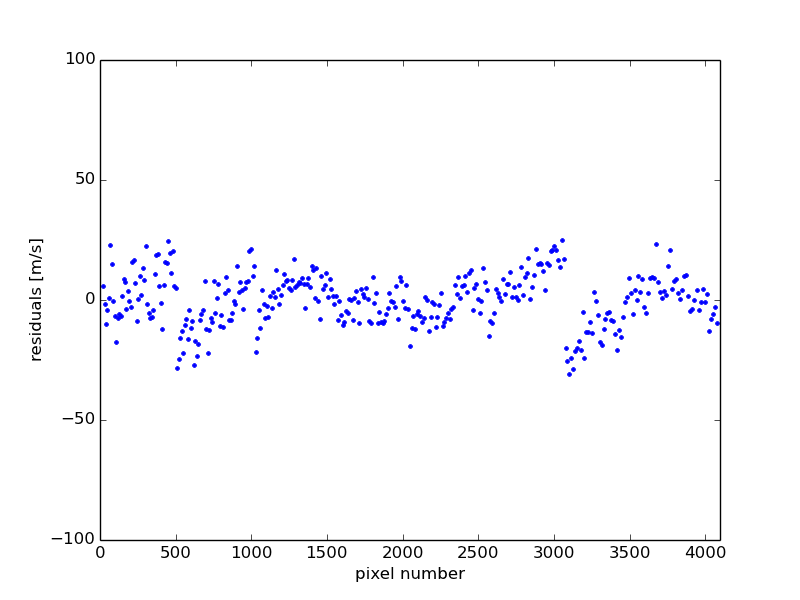}\label{fig:laser_without_gap_ordre_42}}
\hspace*{0.02\textwidth}
\subfloat[Order~42, with gap correction.]{\includegraphics[width=0.49\textwidth, keepaspectratio]{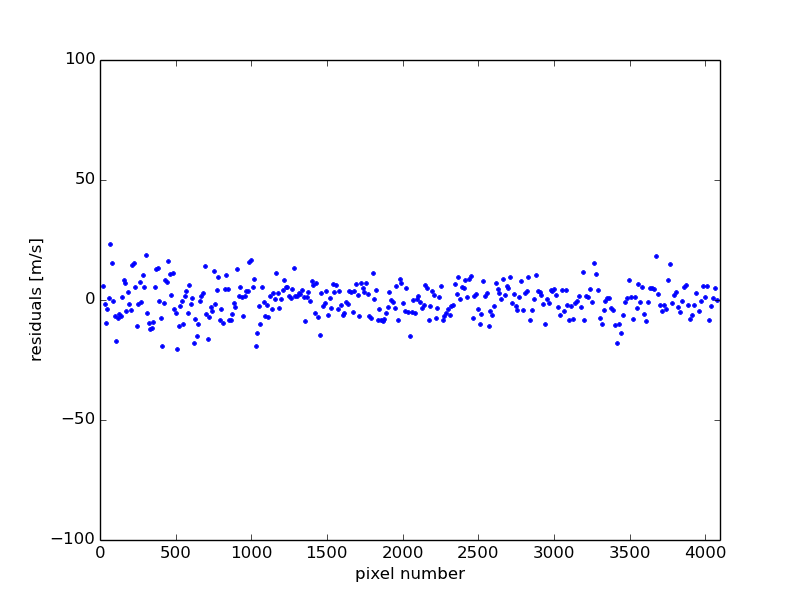}\label{fig:laser_with_gap_ordre_42}}

\subfloat[Order~57, without gap correction.]{\includegraphics[width=0.49\textwidth, keepaspectratio]{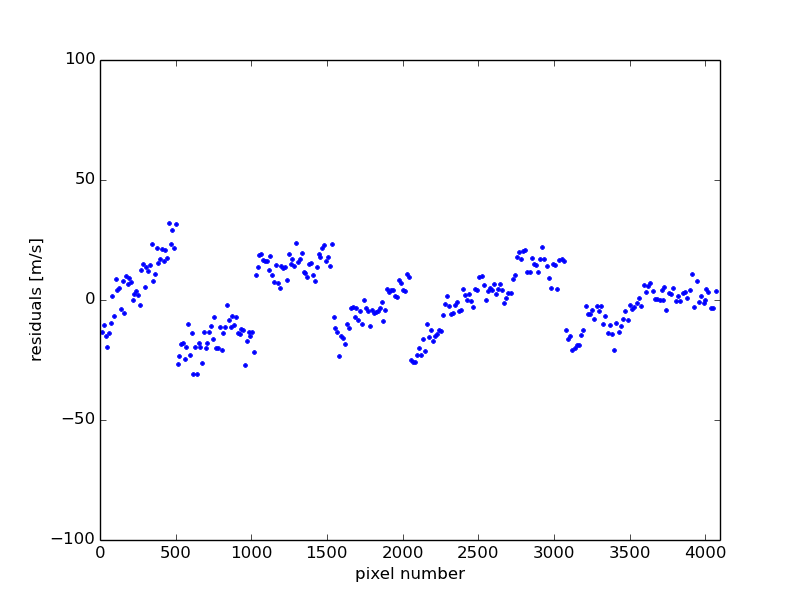}\label{fig:laser_without_gap_ordre_57}}
\hspace*{0.02\textwidth}
\subfloat[Order~57, with gap correction.]{\includegraphics[width=0.49\textwidth, keepaspectratio]{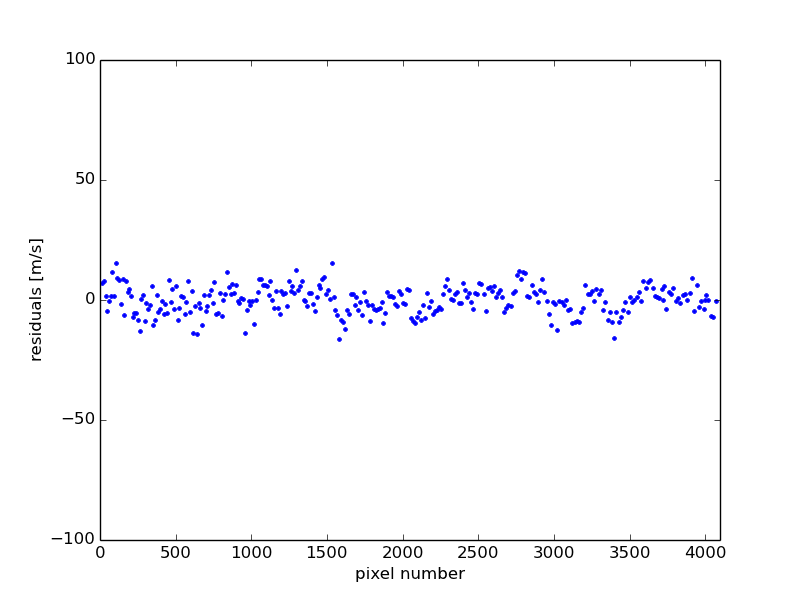}\label{fig:laser_with_gap_ordre_57}}

\caption{Comparison of the residuals of laser frequency comb wavelength solutions, with or without correction of the gaps, for orders~25, 42 and 57.}
\label{fig:laser_with_or_without_gap}
\end{figure*}

The position of the lines were first fitted in pixels, i.e. without correcting the gaps, and then the wavelength solutions were computed. Clear jumps were visible in the residuals. The same was then done using a corrected position of the lines, corresponding to the pixel-position corrected by the gaps we estimated before. If these corrected positions correspond to the actual physical position of the lines, we should not see any jump in the residuals. As can be seen on the spectral orders shown in Figure~\ref{fig:laser_with_or_without_gap}, by doing this correction, we do not see any noticeable jump, what indicates that the estimated sizes of the gaps are correct within the residual dispersion of about \SI{7}{\m\per\s}.

This low residual dispersion of the laser solutions, about~\SI{7}{\m\per\s}, i.e. about \SI{9}{\milli\px}, indicates that the intrinsic dispersion due to instrumental effects is not larger than this value. This rules out this possibility as a major explanation of the larger dispersion (\textasciitilde 50--\SI{70}{\m\per\s}) observed with thorium frames. It reinforces the hints that the large scatter seen on thorium solutions is mainly due to intrinsic features of the thorium spectrum.

\subsection{Comparison of the thorium and laser solutions}
     \label{sec:Comparison of the thorium and laser solutions}

\begin{figure*}[htbp!]
\centering

\subfloat[Order~25.]{\includegraphics[angle=270,width=0.49\textwidth, keepaspectratio]
{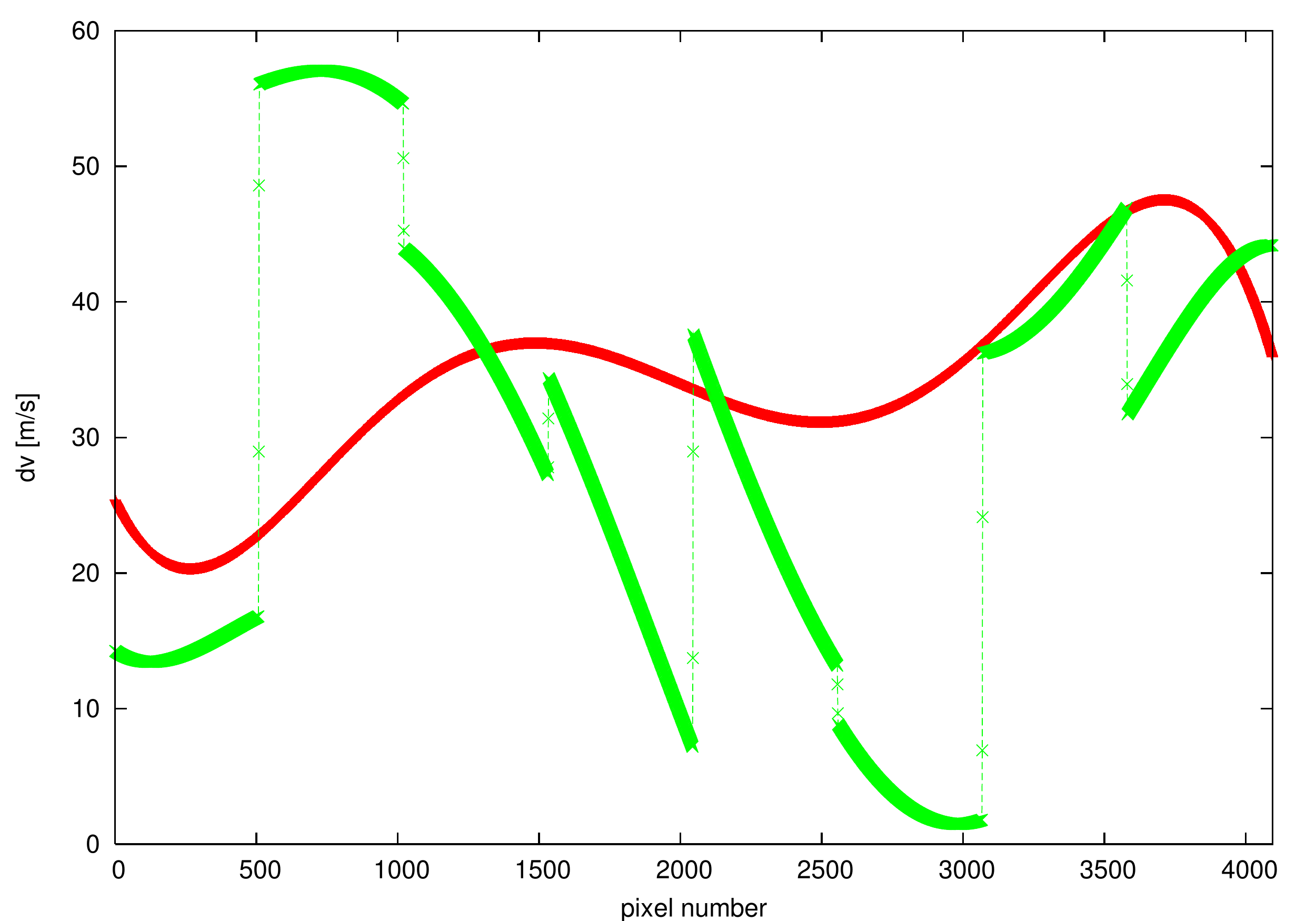}\label{fig:laser_without_gap_ordre_25}}

\subfloat[Order~42.]{\includegraphics[angle=270,width=0.49\textwidth, keepaspectratio]{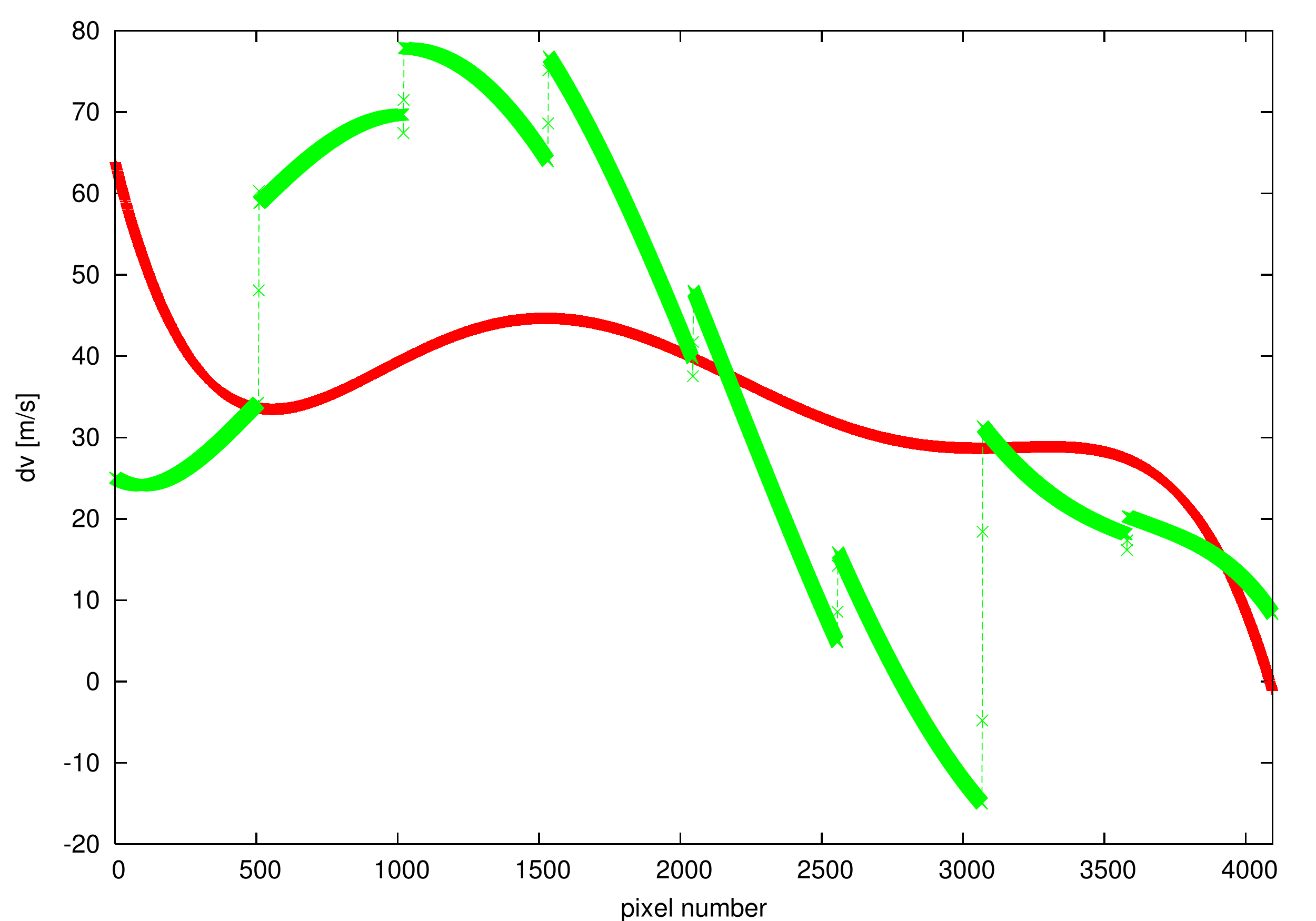}\label{fig:laser_without_gap_ordre_42}}

\subfloat[Order~57.]{\includegraphics[angle=270,width=0.49\textwidth, keepaspectratio]{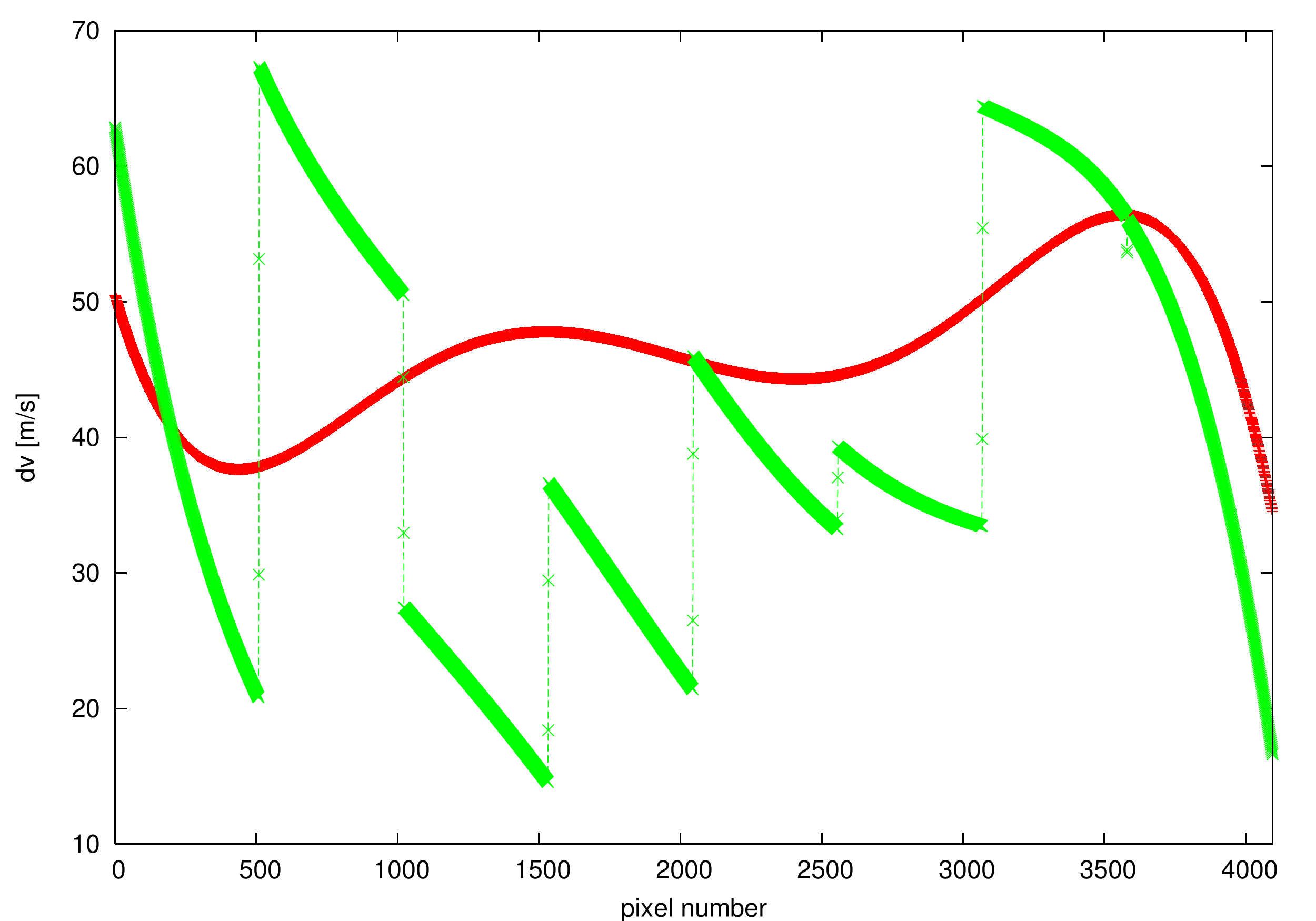}\label{fig:laser_without_gap_ordre_57}}

\caption{Difference, for the same orders as Figure~\ref{fig:laser_with_or_without_gap}, of the wavelength solution from the laser frequency comb and that from the thorium, i.e $c_0\times(\lambda_\mathrm{Th}-\lambda_\mathrm{LFC})/\lambda_\mathrm{LFC}$, where $c_0$ is the speed of light in vacuum, for both the standard (green) and new (red) thorium solutions. The new thorium wavelength solutions match the laser frequency comb solutions much more closely. Especially, we can see that there are ``jumps'' on the green curves, while the red curves are continuous. Note that the laser solutions are 5th-order polynomials while the thorium solutions are 3rd-order polynomials.}
\label{fig:thorium_vs_laser}
\end{figure*}

We further compare the laser and thorium wavelength solutions. In Figure~\ref{fig:thorium_vs_laser}, we plot the difference between the wavelength solutions from the laser and the thorium, for both the uncorrected (standard) and corrected (new) thorium solutions. As can be seen, the difference in much flatter in the case of the new version of the data reduction software (hereafter DRS) than with the standard one. In the case of the standard reduction, we can also clearly see the jumps where the gaps are, while these are not present with the new reduction.

One can however notice that the new wavelength solution has an average offset of about \SI{43}{\m\per\s} with respect to the laser solution, while the standard thorium solution has an offset of about \SI{49}{\m\per\s}. The offset seems to be larger for longer wavelengths. As we explained in Section~\ref{subsec:Laser frames}, the theoretical wavelengths of the laser rfequency comb (LFC) respect the formula $\nu_n = \nu_0 + n \cdot \Delta\nu$, where $\Delta\nu$ (\SI{18}{GHz} in our case) is known at the Hz level and $\nu_0$ was set at \SI{288.0598}{THz} (LFC team, private communication). We note that a fixed error $\Delta \nu_0$ on $\nu_0$ would translate into an error in RV proportional to the wavelength: $\Delta v=\Delta \nu_0 \cdot \lambda$. An error $\Delta \nu_0 = \SI{100}{MHz}$ would translate into an offset $\Delta v = 40$, 50, and \SI{60}{\m\per\s} for a wavelength $\lambda = 4000$, 5000, and \SI{6000}{\AA} respectively. This is close to what we observe. The LFC frames we are using were obtained during the LFC commissioning, during which many changes to the comb properties occurred. In particular, $\nu_0$ was varied by a few hundred MHz over the course of the commissioning. A mistake in the reported $\nu_0$ for the specific frames we are using is thus a plausible explanation.

This is further strengthened by the fact that \cite{Wilken+2010} did not observe any offset between LFC and thorium on the single spectral order that was available at that time. While the thorium solutions have not changed in time, the anchor frequency $\nu_0$ of the LFC has over the different commissioning periods. Unfortunately, we are forced to use the 2015 LFC data because these are the ones providing by far the largest spectral coverage.

We believe that the other possible causes for the offset are less likely than a $\nu_0$ problem. Thorium wavelengths from \citealias{RNS14} have been remeasured several times by the NIST teams over the years, yielding consistent results. Furthermore, any systematic effects in fitting LFC and thorium line positions should also have appeared in \cite{Wilken+2010}, which is not the case. In conclusion, a wrong $\nu_0$ is the most likely culprit for the RV offset we observe. This is unfortunate but we have no means to investigate this further. We stress that, if our hypothesis is correct, the thorium solutions will be correct also in absolute terms.

We can note that, if we compare the two thorium wavelength solutions, we have a mean offset of $c \times \frac{\lambda_\mathrm{new}-\lambda_\mathrm{old}}{\lambda_\mathrm{laser}} \approx \SI{-6}{\m\per\s}$. This is fully compatible with the correction factor $\kappa=0.22\times 10^{-7}$ applied by \citealias{RNS14} on \citealias{PE83}'s wavenumbers, that corresponds to an offset of $c \times \frac{\lambda_\mathrm{RNS14}-\lambda_\mathrm{PE83}}{\lambda_\mathrm{RNS14}} = \SI{-6.6}{\m\per\s}$. This has to be expected since the recomputation of the wavelengths by \citealias{LP07} and by us are fit, for the absolute values, on \citealias{PE83} and \citealias{RNS14} respectively.

\subsection{Global results on stars}
     \label{sec:Global results}

\begin{figure}
   \subfloat[Before the change of fibres (2003-2015).]{
      \includegraphics[width=1\linewidth]{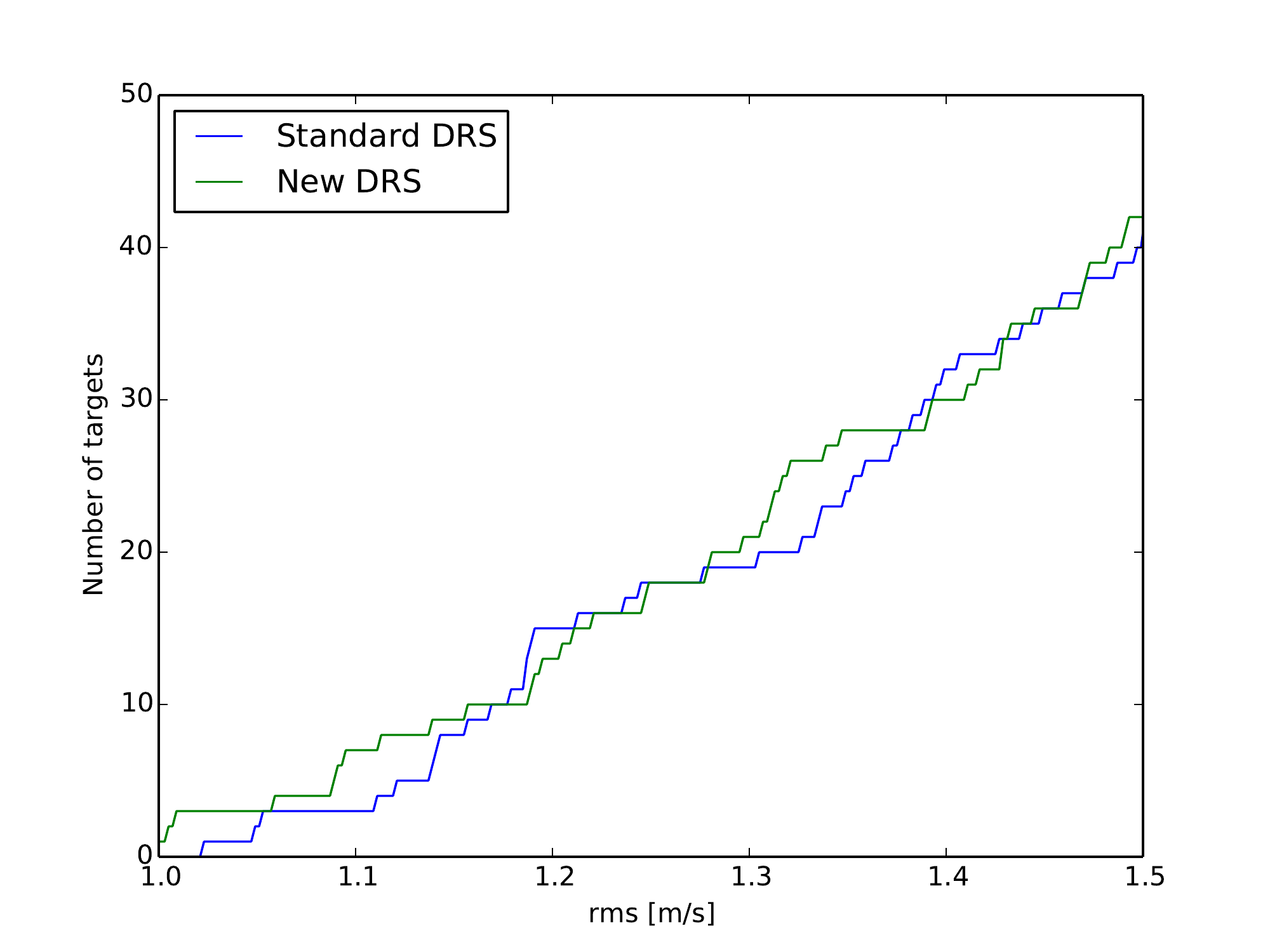}
      \label{fig:histo cumul harps pre 2015}
   } \\
   \subfloat[After the change of fibres (2015-2017).]{
      \includegraphics[width=1\linewidth]{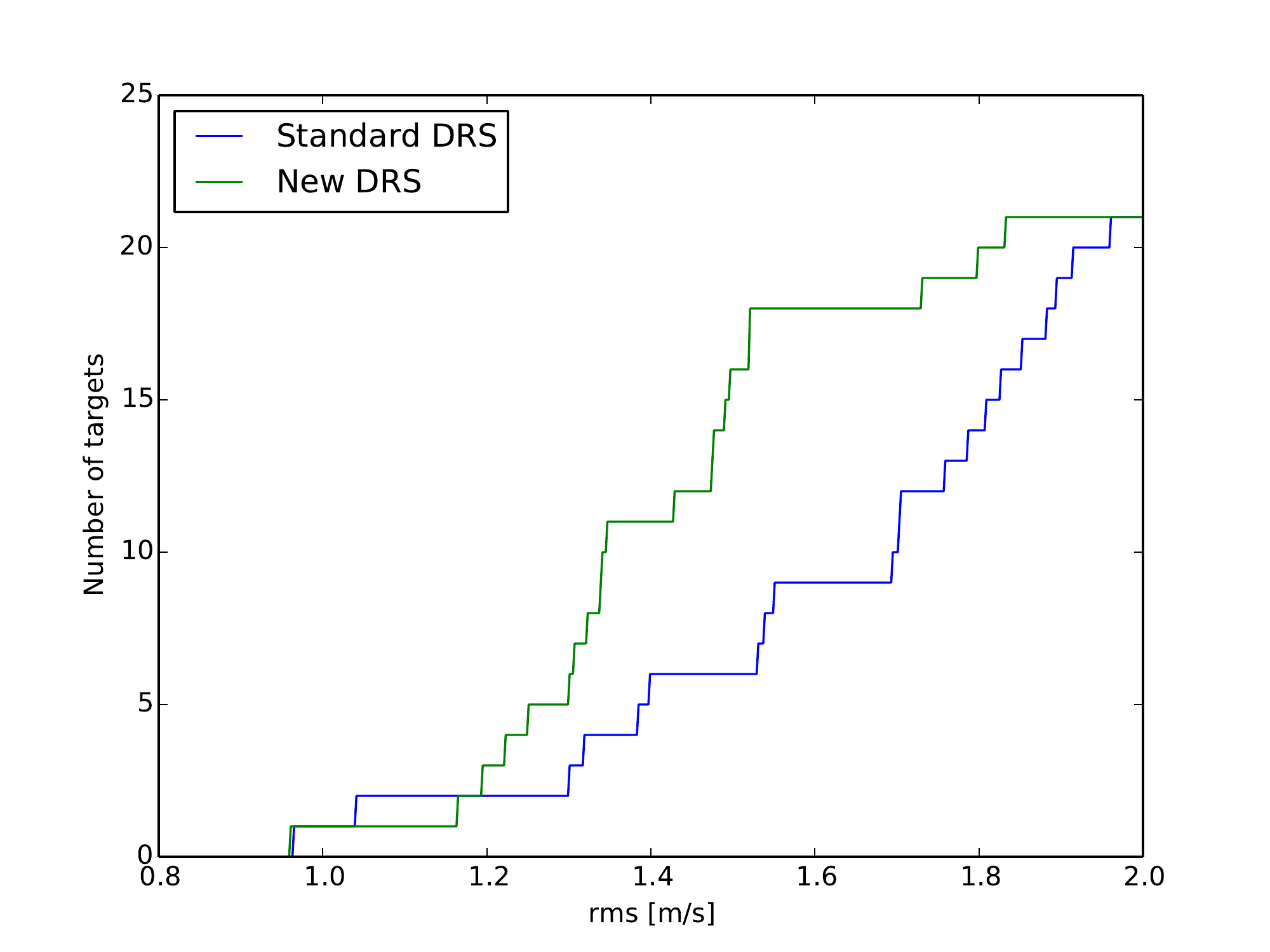}
      \label{fig:histo cumul harps post 2015}
   } \\
   \caption{Cumulative histogram of the rms of the HARPS RVs of all the objects observed at least 10~times during the period by HARPS that have a radial velocity rms below \SI{2}{\m\per\s} with both the standard (blue) and the new (green) versions of the DRS.}
   \label{fig:histo cumul harps}
\end{figure}

A global test to check whether our implementation improved the velocities is to check whether the rms of the most stable stars has decreased with the new DRS with respect to the standard DRS. This comparison is done on Fig.~\ref{fig:histo cumul harps} for the observations before the change of the fibres (observations until May~2015 included, Fig.~\ref{fig:histo cumul harps pre 2015}) and after the change of fibres (since June~2015, Fig.~\ref{fig:histo cumul harps post 2015}). In both cases, we took the subset of objects observed at least 10~times during the period by HARPS that have a radial velocity rms below \SI{2}{\m\per\s} with both the standard and the new versions of the DRS. We therefore have a set of 124~stars for before the change of fibres and 21~stars for after the change of fibres.

\begin{figure}
    \centering
   \includegraphics[scale=0.45]{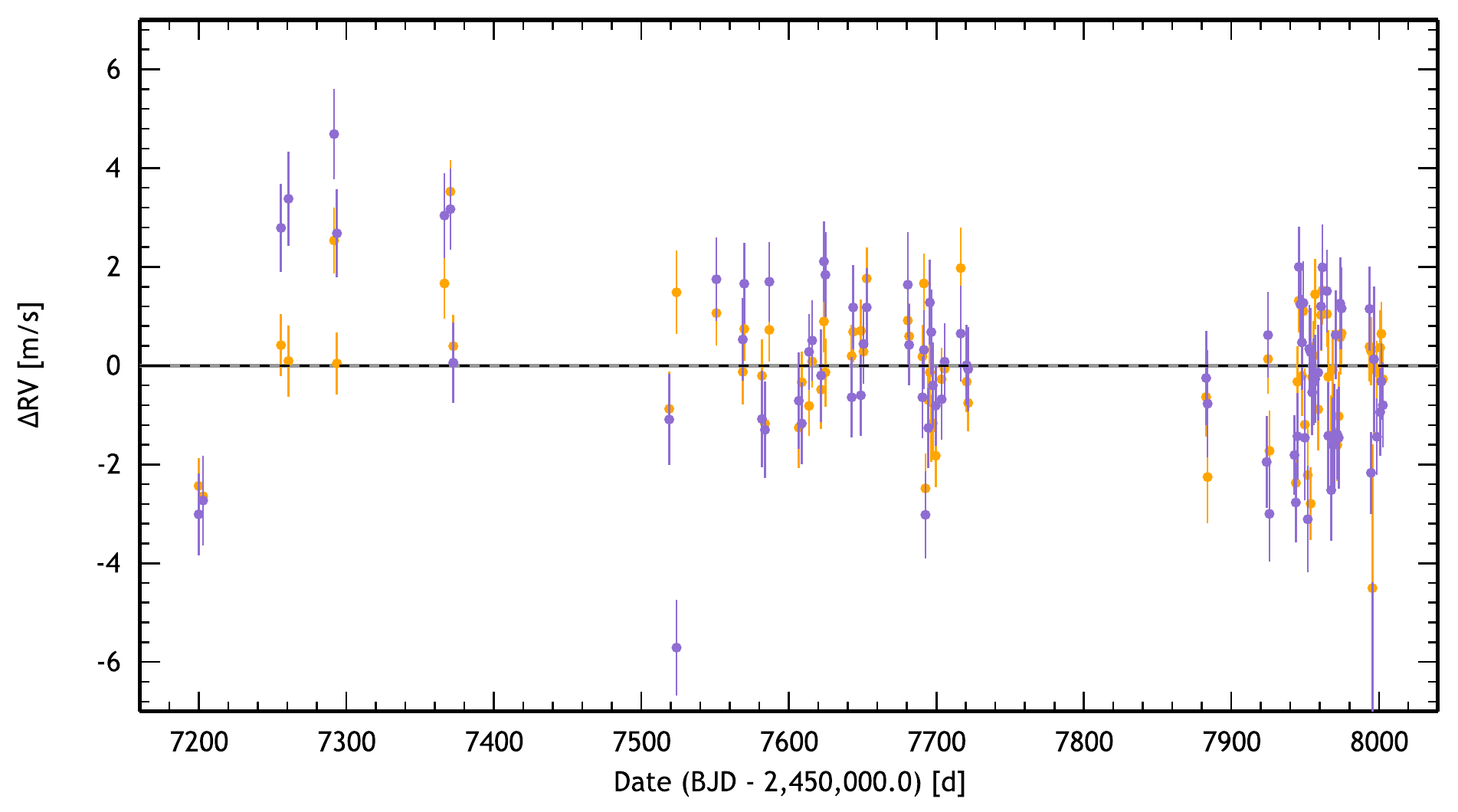}
   \caption{HARPS RVs of HD~220507 after the change of fibres (82~points, $\Delta t = 803~\mathrm{days}$). The rms of the data is \SI{1.73}{\m\per\s} with the standard version of the DRS (purple) while it reduces to \SI{1.23}{\m\per\s} with the new version of the DRS (orange). No significant signal is observed for this star.}
   \label{fig:HD220507_postu}
\end{figure}

For the observations before May~2015, we can see from the cumulative histograms that the new DRS provides a clear improvement for the most stable stars. Indeed, the 10 stars with the lowest RV dispersions with both reductions all have their rms improved with the new DRS. For the observations after the change of the fibres, we can see that the rms are globally significantly improved. The 20 most stable stars all have their rms improved (except one), in many cases by a significant margin. An example of this improvement is shown on Figure~\ref{fig:HD220507_postu}, where we can see the RVs of star HD~220507 after the change of fibre for both the standard and the new versions of the DRS. For this star, where no significant signal is observed, the rms on 82~points covering 803~days after the change of fibres evolves from \SI{1.70}{\m\per\s} to \SI{1.23}{\m\per\s}. This improvement is especially due to the recomputation of the thorium wavelengths for the post-upgrade HARPS as explained in Section~\ref{sec:New wavelength solutions}.

\subsection{Analysis of individual systems}
     \label{sec:Analysis of individual systems}

Given the consistency of the previous results, we further test the new reduction on a set of well-observed stars followed over many years with HARPS.

All the data obtained with HARPS for these stars, spanning up to nearly 5200~days (more than \SI{14}{years}) in total, were reduced with the new pipeline taking into account the block stitching and the newly-defined thorium wavelengths. These results were compared to the ones obtained with the standard pipeline, that did not take into account the gaps and used the wavelengths as defined in \citetalias{LP07}. Notably, we compared the periodograms, to check that the results concerning the known planetary signals and the long-term trend remained basically unchanged, and that at the same time the previously-present 1-year-period signal and its aliases had disappeared, or at least significantly decreased. The periodograms and Keplerian fits presented below were obtained thanks to the Data \& Analysis Center for Exoplanets\footnote{\url{https://dace.unige.ch}} (DACE; see \citealt{Buchschacher+2015}). Keplerian model initial conditions are computed using the formalism described in \cite{Delisle+2016}.

The systems that we analysed are \object{HD~10700}, \object{HD~20794}, \object{HD~69830}, \object{HD~1461}, \object{HD~154088}, \object{HD~31527}, and \object{HD~40307}.

For all the systems, a noise level of \SI{0.75}{\m\per\s} was added quadratically to the DRS-computed RV uncertainties to take into account stellar jitter and the typical instrumental noise floor seen in HARPS data. In addition to the signals fitted, we allow an offset to be adjusted between the two sets of data before and after the change of fibres. All the false-alarm probabilities (hereafter FAP) given here were computed analytically by DACE following \cite{Baluev2008}.

In Table~\ref{tab:table-rms} are listed the planetary systems that we studied, the number of points that we used, the timespan covered by the data, the model that we fitted, the rms of the data after removing the aforementioned model (global rms, rms before and rms after the change of fibres, with both the standard and the new versions of the DRS), as well as complementary comments on the remaining signals. The detailed model for each system is explained on individuals sections.

\begin{table*}[h]
\centering
\begin{tabular}{|c|c|c|c|r|r|r|r|r|r|l|}
  \hline
   System & Points & $\Delta t$ & Model & \multicolumn{2}{c|}{Global rms} & \multicolumn{2}{c|}{rms before} & \multicolumn{2}{c|}{rms after} & Remaining significant signals \\
  HD \# & & & & std & new & std & new & std & new &  \\
  \hline
  10700 & 561 & 5195 & 0 pl. & 1.23 & 1.12 & 1.13 & 1.06 & 1.89 & 1.43 & No long-$P$ power. \\
             & 471+90 &    &         &         &         &         &         &         &         & (\SI{20.0}{d}, \SI{0.35}{\m\per\s}) $\rightarrow$ (\SI{20.0}{d}, \SI{0.36}{\m\per\s}) \\
             &        &          &         &         &         &         &         &         &         & (\SI{49.5}{d}, \SI{0.42}{\m\per\s}) $\rightarrow$ (\SI{49.6}{d}, \SI{0.40}{\m\per\s}) \\
             &        &          &         &         &         &         &         &         &         & (\SI{142}{d}, \SI{0.46}{\m\per\s}) $\rightarrow$ (\SI{142}{d}, \SI{0.49}{\m\per\s}) \\
             &        &          &         &         &         &         &         &         &         & (\SI{228}{d}, \SI{0.38}{\m\per\s}) $\rightarrow$ (\SI{230}{d}, \SI{0.44}{\m\per\s}) \\
             &        &          &         &         &         &         &         &         &         & (\SI{244}{d}, \SI{0.44}{\m\per\s}) $\rightarrow$ (\SI{243}{d}, \SI{0.37}{\m\per\s}) \\
             &        &          &         &         &         &         &         &         &         & (\SI{362}{d}, \SI{0.53}{\m\per\s}) $\rightarrow$ (\SI{360}{d}, \SI{0.45}{\m\per\s}) \\
             &        &          &         &         &         &         &         &         &         & (\SI{593}{d}, \SI{0.59}{\m\per\s}) $\rightarrow$ (\SI{596}{d}, \SI{0.50}{\m\per\s}) \\
             &        &          &         &         &         &         &         &         &         & (\SI{716}{d}, \SI{0.56}{\m\per\s}) $\rightarrow$ (\SI{723}{d}, \SI{0.54}{\m\per\s}) \\
             &        &          &         &         &         &         &         &         &         & (\SI{913}{d}, \SI{0.57}{\m\per\s}) $\rightarrow$ (\SI{948}{d}, \SI{0.49}{\m\per\s}) \\
             &        &          &         &         &         &         &         &         &         & (\SI{2680}{d}, \SI{0.50}{\m\per\s}) $\rightarrow$ (\SI{2533}{d}, \SI{0.37}{\m\per\s}) \\
             &        &          &         &         &         &         &         &         &         & 20.0 and 49.6~d = g and h of \citetalias{Tuomi+2013}/\citetalias{Feng+2017}.  \\
             &        &          &         &         &         &         &         &         &         & 142 / 596~d = e (163~d) / f (636~d) of \citetalias{Tuomi+2013}/\citetalias{Feng+2017}?  \\
             &        &          &         &         &         &         &         &         &         & No signal at 14, 35, 94/92~d (b, c, d of \citetalias{Tuomi+2013}/\citetalias{Feng+2017}).  \\
  \hline
  20794 & 594 & 4535 & 2 pl. & 1.17 & 1.14 & 1.13 & 1.13 & 1.58 & 1.26 & Small long-$P$ power. \\
                   & 530+64 &  & (\citetalias{PLS+11}) &         &         &         &         &         &         & (232 d, \SI{0.53}{\m\per\s}) $\rightarrow$ (237 d, \SI{0.55}{\m\per\s}); \\
                   &        &  &       &         &         &         &         &         &         & (617 d, \SI{0.57}{\m\per\s}) $\rightarrow$ (628 d, \SI{0.49}{\m\per\s}); \\
                   &        &          &          &         &         &         &         &         &         & (1877 d, \SI{0.40}{\m\per\s}) $\rightarrow$ (1815 d, \SI{0.42}{\m\per\s}). \\
                   &        &          &          &         &         &         &         &         &         & Do not correspond to objects of \citetalias{PLS+11} or \citetalias{FTJ17}. \\
  \hline
  69830 & 257 & 4817 & 3 pl. & 1.28 & 1.26 & 1.27 & 1.26 & 1.60 & 1.40 & Long-$P$ power. \\
             & 249+8 &      & (\citetalias{Lovis+2006}) &      &      &      &      &      &      &  (363 d, \SI{0.68}{\m\per\s}) $\rightarrow$ (363 d, \SI{0.63}{\m\per\s}) \\
             &        &      &       &      &      &      &      &      &      &  (4006 d, \SI{0.58}{\m\per\s}) $\rightarrow$ (4098 d, \SI{0.58}{\m\per\s}): MC \\
  \hline
1461      & 266 & 5073 & 2+MC & 1.68 & 1.64 & 1.65 & 1.60 & 2.54 & 2.84 & No long-$P$ power. \\
              & 258+8 &      & (\citetalias{Diaz+2016}) &         &         &         &         &         &         & (365, 1.27) $\rightarrow$ (366, 1.15). \\
  \hline
  154088 & 189 & 4144 & 1 pl. & 1.78 & 1.74 & 1.78 & 1.74 & 0.70 & 0.03 & Long-$P$ power but no significant signal. \\
               & 187+2 &    & (\citetalias{Mayor+2011}) &         &         &         &         &         &         & \textit{The rms after, based on 2 points, is not meaningful.} \\
               &        &          & 1 + d1 & 1.71 & 1.65 & 1.72 & 1.66 & 0.67 & 0.06 & Very low long-$P$ power. No significant signal. \\
               &            &    & (\citetalias{Lovis+2011}) &         &         &         &         &         &         & \textit{The rms after, based on 2 points, is not meaningful.} \\
  \hline
  31527 & 256 & 5053 & 3 pl. & 1.38 & 1.42 & 1.39 & 1.42 & 1.37 & 1.42 & Low long-$P$ power but no signal with a \\
                   & 245+11 &  & (\citetalias{Mayor+2011}) &        &         &         &         &         &         & $p$-value<10\%. \\
  \hline
  40307 & 242 & 5059 & 4 + d3 & 1.31 & 1.32 & 1.30 & 1.30 & 1.54 & 1.70 & No long-$P$ power. No significant signal. \\
             & 232+10 &     & (\citetalias{Mayor+2009}) &         &         &         &         &         &         & Especially, the signal at 200 d is no more \\
             &        &         & /\citetalias{TAE+2013}  &         &         &         &         &         &         & significant when we fit a cubic drift \\
             &        &         & /\citetalias{Diaz+2016}) &         &         &         &         &         &         & and no signal at 34.6 d (planet e, \citetalias{TAE+2013}) \\
  \hline
\end{tabular}
\caption{Comparison of the standard (std) and the new versions of the DRS. The numbers of points are the total and, below, the number before+after the 2015 upgrade. All these points are night-averaged. The rms columns provide the post-fit RV dispersions, both before and after the fibre upgrade. All the rms are in \si{\m\per\s}. The timespans $\Delta t$ are in days. Arrows mean ``std value'' $\rightarrow$ ``new value''. ``$n$~pl.'' means $n$ planetary (Keplerian) signals; ``MC'' means ``magnetic cycle'', also fitted by a Keplerian signal unless stated otherwise. References: \citetalias{Diaz+2016} = \cite{Diaz+2016}; \citetalias{FTJ17} = \cite{FTJ17}; \citetalias{Feng+2017} = \cite{Feng+2017}; \citetalias{Lovis+2006} = \cite{Lovis+2006}; \citetalias{Lovis+2011} = \cite{Lovis+2011}; \citetalias{Mayor+2009} = \cite{Mayor+2009}; \citetalias{Mayor+2011} = \cite{Mayor+2011}; \citetalias{PLS+11} = \cite{PLS+11}; \citetalias{TAE+2013} = \cite{TAE+2013}; \citetalias{Tuomi+2013} = \cite{Tuomi+2013}.}
     \label{tab:table-rms}
\end{table*}

\subsubsection{HD~10700}
     \label{subsec:HD 10700}

HD~10700 (Tau~Ceti) is a G8{V} star located 3.7~parsecs away from the Solar System. It is one of the most stable stars in radial velocities with a long-term dispersion of $\sim$\SI{1}{\m\per\s} \citep{PLS+11,Tuomi+2013}. Both with the standard and the new versions of the DRS, we find a lot of peaks in the periodogram with a FAP below 1\%. The 1-year signal becomes slightly less significant with the new version (FAP of 0.49\%) compared to the standard one (FAP of 0.057\%). While the peaks around 600 and 900~days have a very high significance with the standard version, they have a lower one with the new version, while still having a FAP well below 0.1\%. On the contrary, the peak around 140~days becomes significantly stronger. The 20-day signal, that corresponds to planet~g of \cite{Tuomi+2013} and \cite{Feng+2017}, is basically undetected in the periodogram with the standard version while it becomes marginally significant with the new one. The signal at 49.6~days, that corresponds to planet~h, is detected in both cases with a FAP below 1\%. There seems to be a signal around 230-240~days, even if its double-peak structure makes it unclear, especially given that one of the two peaks is stronger with one version while the other peak is stronger with the other version. A marginally-significant peak is detected around 720~days with the standard version while it is not significant at all with the new version. Finally, the long-period signal around 2700~days, that was very significant with the standard version, becomes insignificant with the new one. While two of these signals (20.0 and 49.6~days) seem to correspond to signals detected by \cite{Tuomi+2013} and \cite{Feng+2017}, the others do not seem to correspond. We may tentatively associate the signal around 600~days to their planet~f and the signal around 142~days to planet~e, but the periods are quite different and this correspondence looks doubtful. Concerning the other planets (b, c and d), no signal is detected at the corresponding announced periods (respectively 14, 35 and 94/92~days).

As a conclusion, we urge caution in the interpretation of the many weak RV signals seen in HD~10700, as we demonstrate that their amplitude and significance depend on the pipeline version used and the data analysis method.

\subsubsection{HD~20794}
     \label{subsec:HD 20794}

HD~20794 is a G6{V} star located 6.0~parsecs away from the Solar System. Two signals, corresponding to planets HD~20794~b and HD~20794~d respectively \citep{PLS+11}, are detected in our periodogram analysis, at \SI[separate-uncertainty = true]{18.317\pm0.005}{d} and \SI[separate-uncertainty = true]{89.65\pm0.10}{d} with the standard DRS and at \SI[separate-uncertainty = true]{18.320\pm0.005}{d} and \SI[separate-uncertainty = true]{89.76\pm0.12}{d} with the new DRS. For the amplitudes of these signals, we find respectively $(0.67\pm0.07)$ and \SI[separate-uncertainty = true]{0.71\pm0.07}{\m\per\s} for HD~20794~b, and $(0.74\pm0.09)$ and \SI[separate-uncertainty = true]{0.57\pm0.07}{\m\per\s} for HD~20794~d. While the difference is within 1~$\sigma$ for planet b, it is more noticeable for planet d, 22\% and nearly 2~$\sigma$ lower with the new DRS than with the standard one. The fitted eccentricities are also quite lower with the new DRS, changing from $0.18\pm0.10$ (b) and $0.45\pm0.09$ (d) with the standard version of the DRS to $0.11\pm0.10$ (b) and $0.29\pm0.11$ (d). The other candidate planets, HD~20794~g, c, e and f at respectively 12, 43, 147 and 330~days \citep{PLS+11,FTJ17}, are not found. However, we see signals around 235, 620 and 1900~days. The first two of these signals may correspond to those around 250 and 680~days found by \cite{FTJ17}, that they interpret respectively as a yearly alias of their signal at 147~days and a harmonic of their signal at 330~days. Given that we do not detect the signals at 147 and 330~days, the actual situation may be the opposite. Again, caution is in order when interpreting the nature of these additional signals, whose properties vary according to the pipeline version used. A full study is beyond the scope of this paper.

\subsubsection{HD~69830}
     \label{subsec:HD 69830}

HD~69830 is a K0{V} star located 12.6~parsecs away from the Solar System. The three planets announced in \cite{Lovis+2006}, with periods of 8.67, 31.6 and about 200~days, are recovered. After fitting these three planets, two peaks remain visible below the significance threshold: one around one year and a long-period one around 4000~days, that may correspond to the magnetic cycle of the star. The one-year peak is slightly less significant with the new DRS while the long-period peak is slightly more significant. Since the long-period signal is not significant, we did not fit it here. However, we checked and found that fitting it ($P\approx\SI{3680}{d}, K\approx\SI{1.1}{\m\per\s}, e\approx0.8$) does not change much the values we obtain for the planets, all remaining well within one sigma. For the innermost planet, we obtain a period of \SI[separate-uncertainty = true]{8.6682\pm0.0004}{days}, an amplitude of \SI[separate-uncertainty = true]{3.47\pm0.11}{\m\per\s} and an eccentricity of \SI[separate-uncertainty = true]{0.07\pm0.03}{}. For the second planet, the period is \SI[separate-uncertainty = true]{31.601\pm0.006}{days}, the amplitude \SI[separate-uncertainty = true]{2.66\pm0.13}{\m\per\s} and the eccentricity \SI[separate-uncertainty = true]{0.06\pm0.05}{}. Finally, the outermost planet is found to have a period of \SI[separate-uncertainty = true]{200.32\pm0.36}{days}, an amplitude of \SI[separate-uncertainty = true]{1.85\pm0.12}{\m\per\s} and an eccentricity of \SI[separate-uncertainty = true]{0.12\pm0.06}{} with the standard version of the DRS while we find a period of \SI[separate-uncertainty = true]{200.57\pm0.37}{days}, an amplitude of \SI[separate-uncertainty = true]{1.77\pm0.12}{\m\per\s} and an eccentricity of \SI[separate-uncertainty = true]{0.10\pm0.06}{} with the new version.

\subsubsection{HD~1461}
     \label{subsec:HD 1461}

HD~1461 is a G0{V} star located 23.4~parsecs away from the Solar System. \cite{Diaz+2016} announced two planets (periods of 5.77 and \SI{13.5}{days}) around this star as well as a magnetic cycle (period of \SI{3500}{days}). We do find these three signals with both the standard and the new version of the DRS. Without fitting the magnetic cycle, we find a signal at 37.3~days with both methods, but while its FAP is slightly below 1\% with the standard version of the DRS, it is slightly above with the new version. This signal disappears once we fit the magnetic cycle. Both with and without fitting the magnetic cycle, we find a significant signal around 1~year, even if we can notice that the amplitude is a bit smaller with the new version of the DRS.

All the parameter values obtained for this system from the old and new versions of the DRS are well within $1~\sigma$ of each other, so here we give only the values from the new DRS. The fit of the magnetic cycle gives a period of \SI[separate-uncertainty = true]{3582\pm80}{days}, an amplitude of \SI[separate-uncertainty = true]{3.28\pm0.28}{\m\per\s} and an eccentricity of \SI[separate-uncertainty = true]{0.51\pm0.07}{}. Concerning the innermost planet, we find a period of \SI[separate-uncertainty = true]{5.7715\pm0.0004}{days}, an amplitude of \SI[separate-uncertainty = true]{2.26\pm0.15}{\m\per\s} and an eccentricity of \SI[separate-uncertainty = true]{0.11\pm0.07}{}. For the second planet, we get a period of \SI[separate-uncertainty = true]{13.506\pm0.003}{days}, an amplitude of \SI[separate-uncertainty = true]{1.60\pm0.16}{\m\per\s} and an eccentricity of \SI[separate-uncertainty = true]{0.15\pm0.10}{}.

\subsubsection{HD~154088}
     \label{subsec:HD 154088}

HD~154088 is a K0{IV} star located 17.8~parsecs away from the Solar System. The single planet announced by \cite{Mayor+2011} at a period of 18.6 days around this star is easily recovered. No other significant signal is found in the HARPS RV data of this star, even if we can notice the presence of some low power at long periods in the periodogram. We therefore decided to model these data both with a single Keplerian and with a Keplerian plus a linear drift. In both cases, the values from the old and the new DRS are all compatible within $1~\sigma$ so, as before, we give only the values from the new DRS.

From the model with only a Keplerian, we find a period of \SI[separate-uncertainty = true]{18.556\pm0.009}{days}, nearly $2~\sigma$ lower than the \SI[separate-uncertainty = true]{18.596\pm0.021}{days} in \cite{Mayor+2011}. The amplitude is found to be \SI[separate-uncertainty = true]{1.68\pm0.19}{\m\per\s}, slightly lower than but well within $1~\sigma$ of the \SI[separate-uncertainty = true]{1.78\pm0.31}{\m\per\s} of \cite{Mayor+2011}. We find an eccentricity of \SI[separate-uncertainty = true]{0.10\pm0.11}{}, noticeably lower than the \SI[separate-uncertainty = true]{0.38\pm0.15}{} in \cite{Mayor+2011}.

According to \cite{Lovis+2011}, a long-term magnetic cycle is present in this star, but its period is very unconstrained, given as $8000^{+\infty}_{-6487}$~days in this paper. Here, we can clearly notice a trend in the data. Fitting a linear drift significantly reduces the dispersion of the residuals. This drift has an amplitude of \SI[separate-uncertainty = true]{0.33\pm0.07}{\m\per\s\per\yr}. In this case, we find for the planet a period of \SI[separate-uncertainty = true]{18.568\pm0.008}{days}, slightly larger than before but still lower than what had found \cite{Mayor+2011}. The amplitude is found to be \SI[separate-uncertainty = true]{1.64\pm0.23}{\m\per\s}, slightly lower than before but well within $1~\sigma$. We find an eccentricity of \SI[separate-uncertainty = true]{0.28\pm0.12}{}, noticeably larger than before but still lower than what had found \cite{Mayor+2011}. Even if its significance is extremely low in all cases, one can notice that the peak around 360~days is smaller with the new DRS than with the standard one.

\subsubsection{HD~31527}
     \label{subsec:HD 31527}

HD~31527 is a G0{V} star located 38.4~parsecs away from the Solar System. \cite{Mayor+2011} announced the detection of three planets around this star, at periods of 16, 51 and 274~days. All three signals are easily recovered with both versions of the DRS and no other significant signal is found in the HARPS RV data of this star. The highest remaining signal is a multi-peak between 300-400~days, but with negligible significance.

Both versions of the DRS give values compatible within $1~\sigma$, except for the amplitudes of planet d, that are nonetheless within less than $2~\sigma$. Therefore, here again, we give only the values from the new DRS. Concerning the planets, HD~31527~b is found to have a period of \SI[separate-uncertainty = true]{16.557\pm0.003}{days}, an amplitude of \SI[separate-uncertainty = true]{2.84\pm0.13}{\m\per\s} and an eccentricity of \SI[separate-uncertainty = true]{0.16\pm0.05}{}. HD~31527~c has a period of \SI[separate-uncertainty = true]{51.20\pm0.03}{days}, an amplitude of \SI[separate-uncertainty = true]{2.70\pm0.14}{\m\per\s} and an eccentricity of \SI[separate-uncertainty = true]{0.08\pm0.05}{}. Finally, HD~31527~d has a period of \SI[separate-uncertainty = true]{274.2\pm1.0}{days}, an amplitude of \SI[separate-uncertainty = true]{1.77\pm0.36}{\m\per\s} and an eccentricity of \SI[separate-uncertainty = true]{0.67\pm0.08}{}. All these values are compatible with those given by \cite{Mayor+2011}, even if we find a noticeably larger value for the eccentricity of planet~d. One can also note that, with 89 more points and nearly twice as long a timespan as in \cite{Mayor+2011}, we have significantly lower uncertainties on the parameters of the outermost planet.

\subsubsection{HD~40307}
     \label{subsec:HD 40307}

HD~40307 is a K3{V} star located 13~parsecs away from the Solar System. \cite{Mayor+2009} discovered three planets (b, c and d) around this star, at periods of 4.3, 9.6 and 20.5~days, that we easily recover. \cite{TAE+2013} announced the detection of three more planets (e, f and g), at periods of 34.6, 51.8 and 197.8~days. \cite{Diaz+2016} confirmed the three planets of \cite{Mayor+2009} as well as the 51.6-day planet of \cite{TAE+2013}. While they find a signal at 205~days that may correspond to \cite{TAE+2013}'s planet~g, they conclude that its significance and interpretation are doubtful. They also find that the six-planet model including the planet at 34.6~days is strongly disfavoured. Here, we recover the three planets of \cite{Mayor+2009} as well as the signal at 51.6~days. As found by \cite{Diaz+2016}, we detect a long-term signal, probably the magnetic cycle of the star, that we fitted by a third-degree polynomial. Once this is done, no significant signal remain. Especially, the signal around 200~days has a very low significance and no signal is found around 34.6~days.

The two DRS give values compatible within $1~\sigma$, so here we also give only the values from the new version. The innermost planet has a period of \SI[separate-uncertainty = true]{4.3114\pm0.0002}{days}, an amplitude of \SI[separate-uncertainty = true]{1.84\pm0.14}{\m\per\s} and an eccentricity of \SI[separate-uncertainty = true]{0.12\pm0.07}{}. The second planet has a period of \SI[separate-uncertainty = true]{9.6210\pm0.0008}{days}, an amplitude of \SI[separate-uncertainty = true]{2.29\pm0.13}{\m\per\s} and an eccentricity of \SI[separate-uncertainty = true]{0.05\pm0.06}{}. The third planet has a period of \SI[separate-uncertainty = true]{20.412\pm0.004}{days}, an amplitude of \SI[separate-uncertainty = true]{2.31\pm0.14}{\m\per\s} and an eccentricity of \SI[separate-uncertainty = true]{0.07\pm0.05}{}. The fourth signal has a period of \SI[separate-uncertainty = true]{51.68\pm0.07}{days}, an amplitude of \SI[separate-uncertainty = true]{0.78\pm0.13}{\m\per\s} and an eccentricity of \SI[separate-uncertainty = true]{0.19\pm0.16}{}. Finally, the drift parameters are \SI[separate-uncertainty = true]{0.60\pm0.07}{\m\per\s\per\yr}, \SI[separate-uncertainty = true]{-0.13\pm0.02}{\m\per\s\per\square\yr} and \SI[separate-uncertainty = true]{-0.02\pm0.00}{\m\per\s\per\cubic\yr}.


\begin{figure*}
\centering

\subfloat[HD~10700 standard, k0d0.]{\includegraphics[width=0.49\textwidth, keepaspectratio]{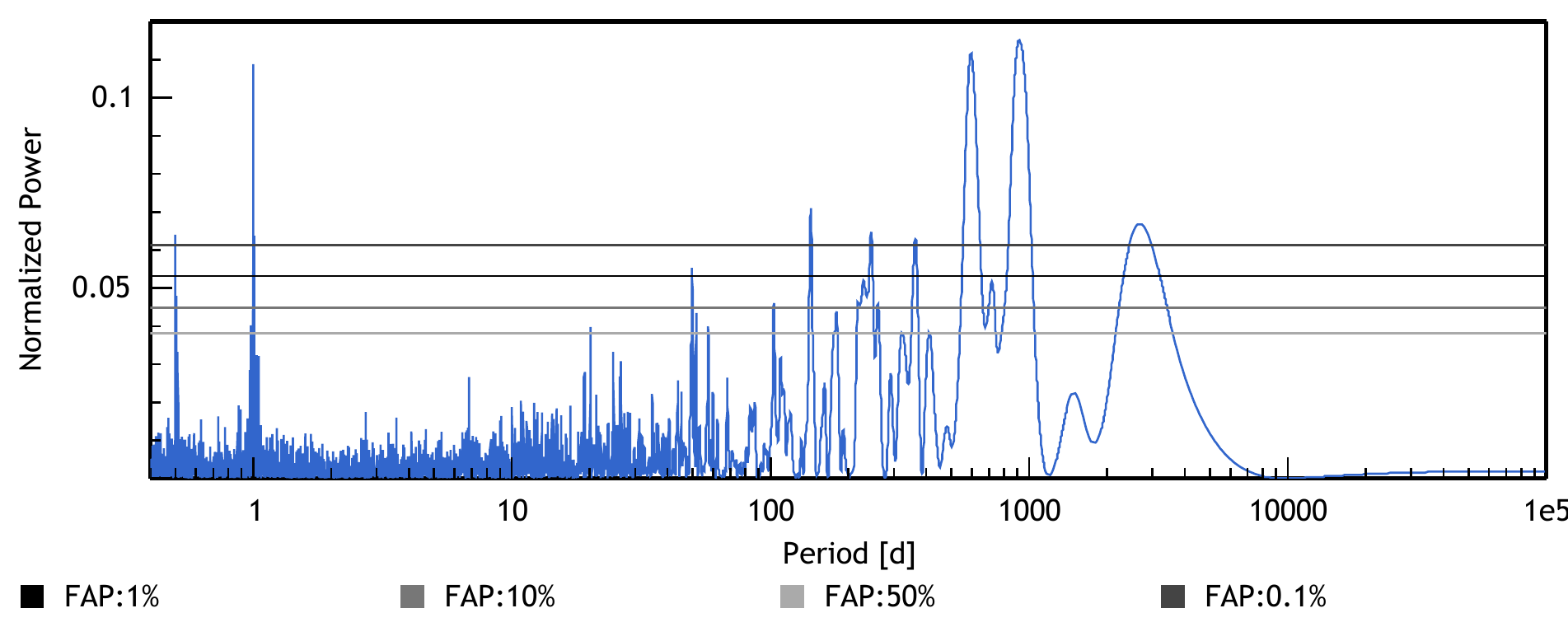}\label{fig:subfig5}}
\hspace*{0.02\textwidth}
\subfloat[HD~10700 new, k0d0.]{\includegraphics[width=0.49\textwidth, keepaspectratio]{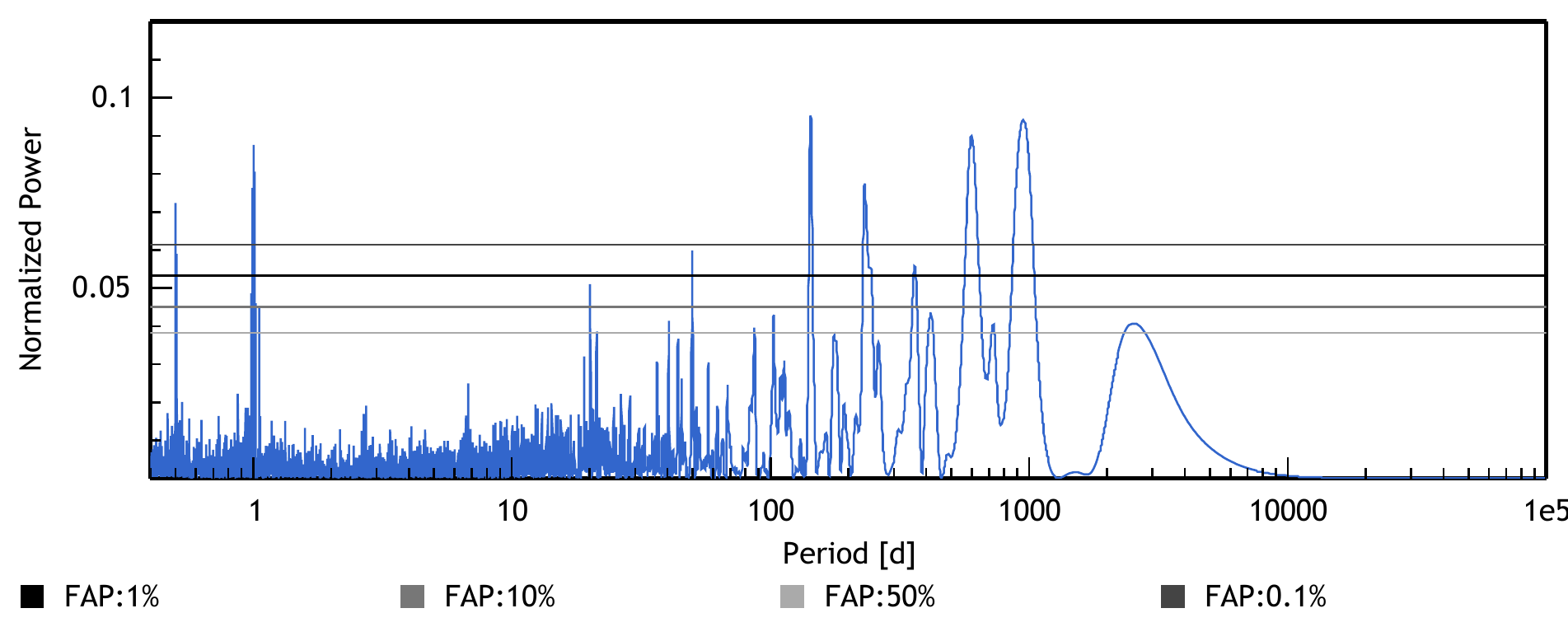}\label{fig:subfig5}}

\caption{Periodograms of HD~10700 without fitting any Keplerian or drift.}
\label{fig:test}
\end{figure*}


\begin{figure*}
\centering

\subfloat[HD~20794 standard, k2d0.]{\includegraphics[width=0.49\textwidth, keepaspectratio]{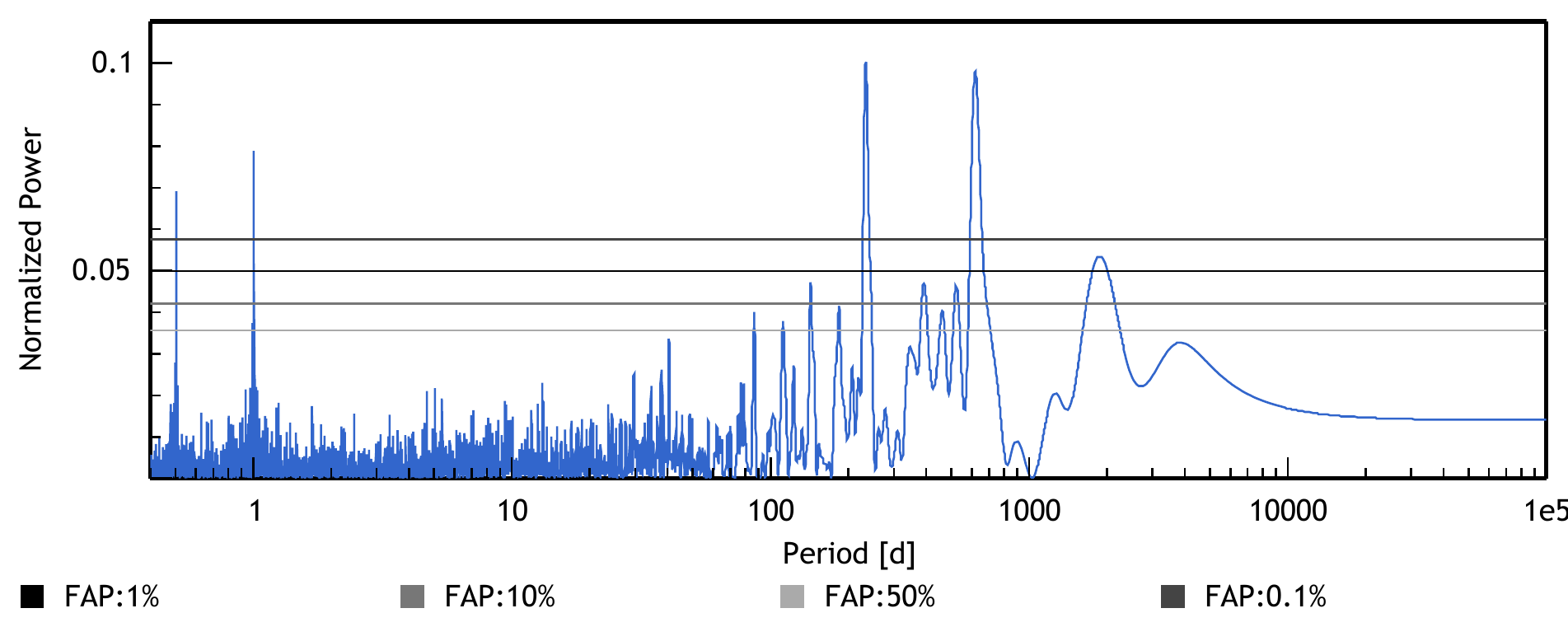}\label{fig:subfig5}}
\hspace*{0.02\textwidth}
\subfloat[HD~20794 new, k2d0.]{\includegraphics[width=0.49\textwidth, keepaspectratio]{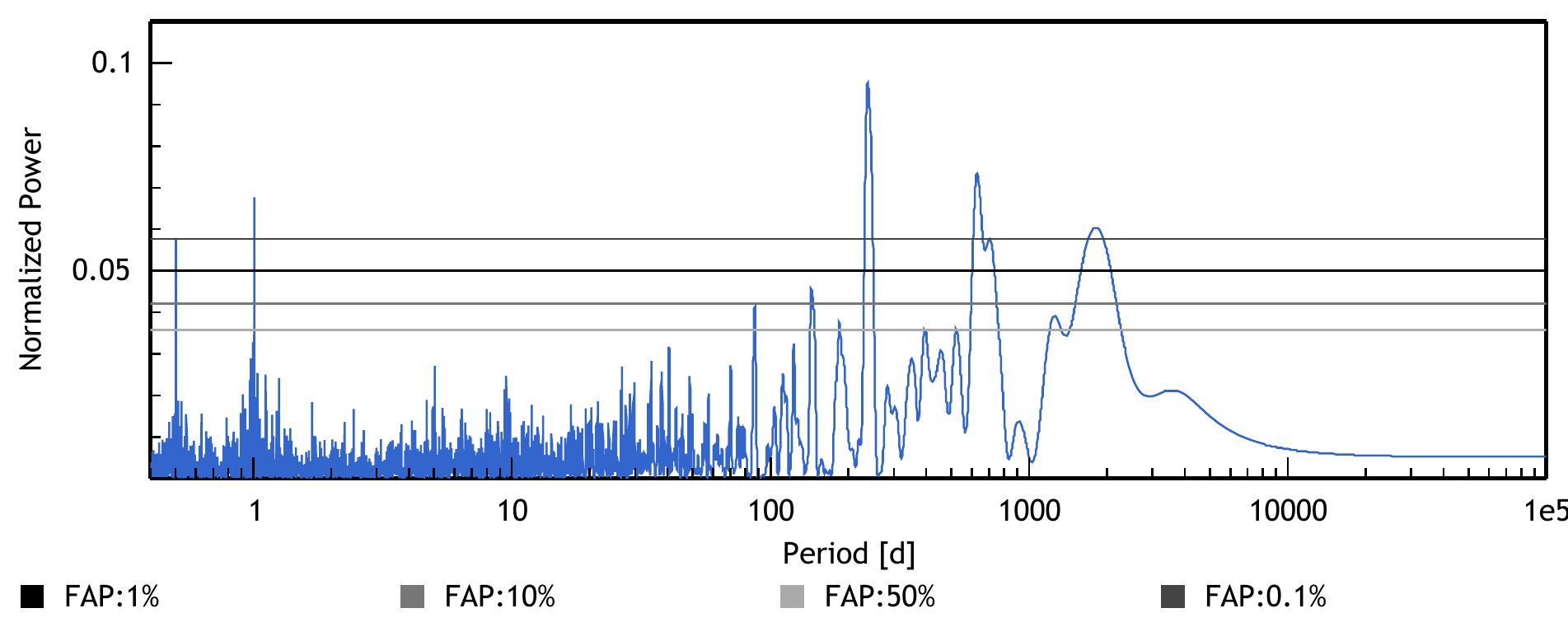}\label{fig:subfig5}}

\caption{Periodograms of HD~20794 after the fit of the two known planets at 18.32 and 89.7~days.}
\label{fig:test}
\end{figure*}


\begin{figure*}
\centering

\subfloat[HD~69830 standard, k3d0.]{\includegraphics[width=0.49\textwidth, keepaspectratio]{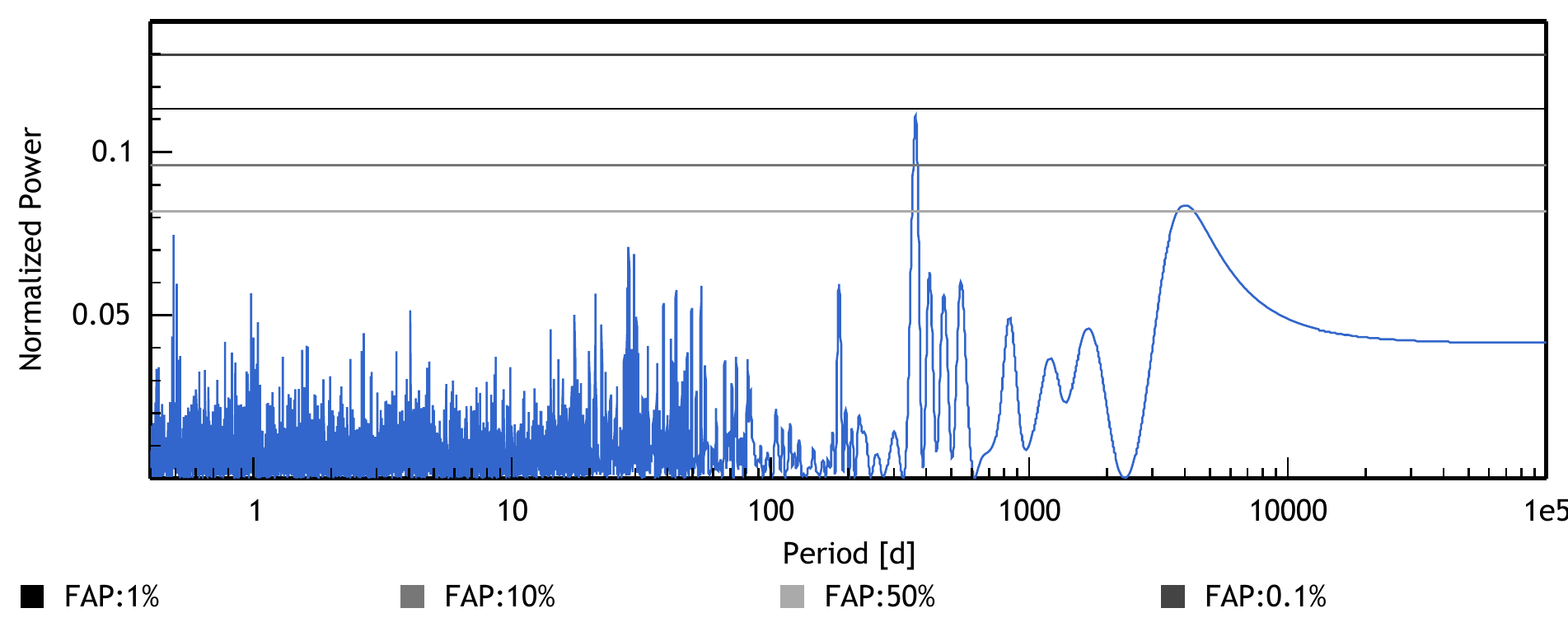}\label{fig:subfig5}}
\hspace*{0.02\textwidth}
\subfloat[HD~69830 new, k3d0.]{\includegraphics[width=0.49\textwidth, keepaspectratio]{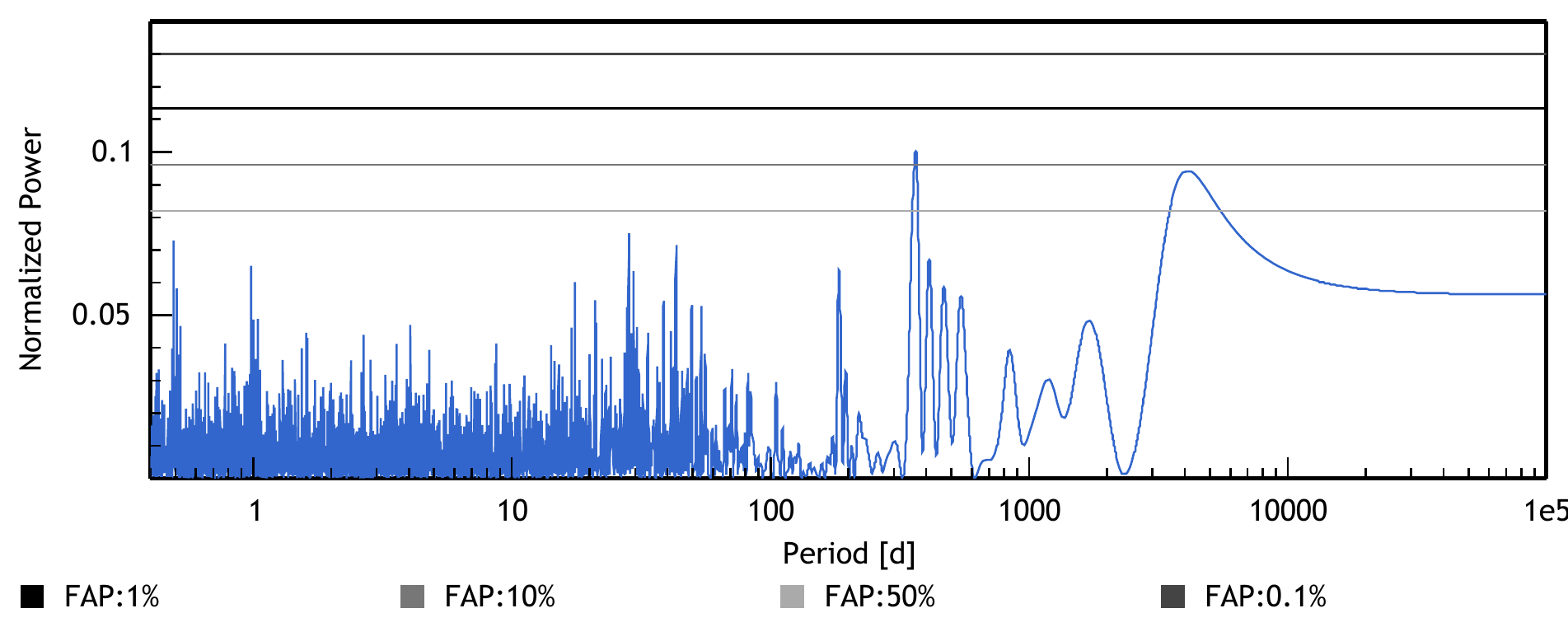}\label{fig:subfig5}}

\caption{Periodograms of HD~69830 after the fit of the three known planets at periods of 8.67, 31.60 and 200~days.}
\label{fig:test}
\end{figure*}


\begin{figure*}
\centering

\subfloat[HD~1461 standard, k3d0.]{\includegraphics[width=0.49\textwidth, keepaspectratio]{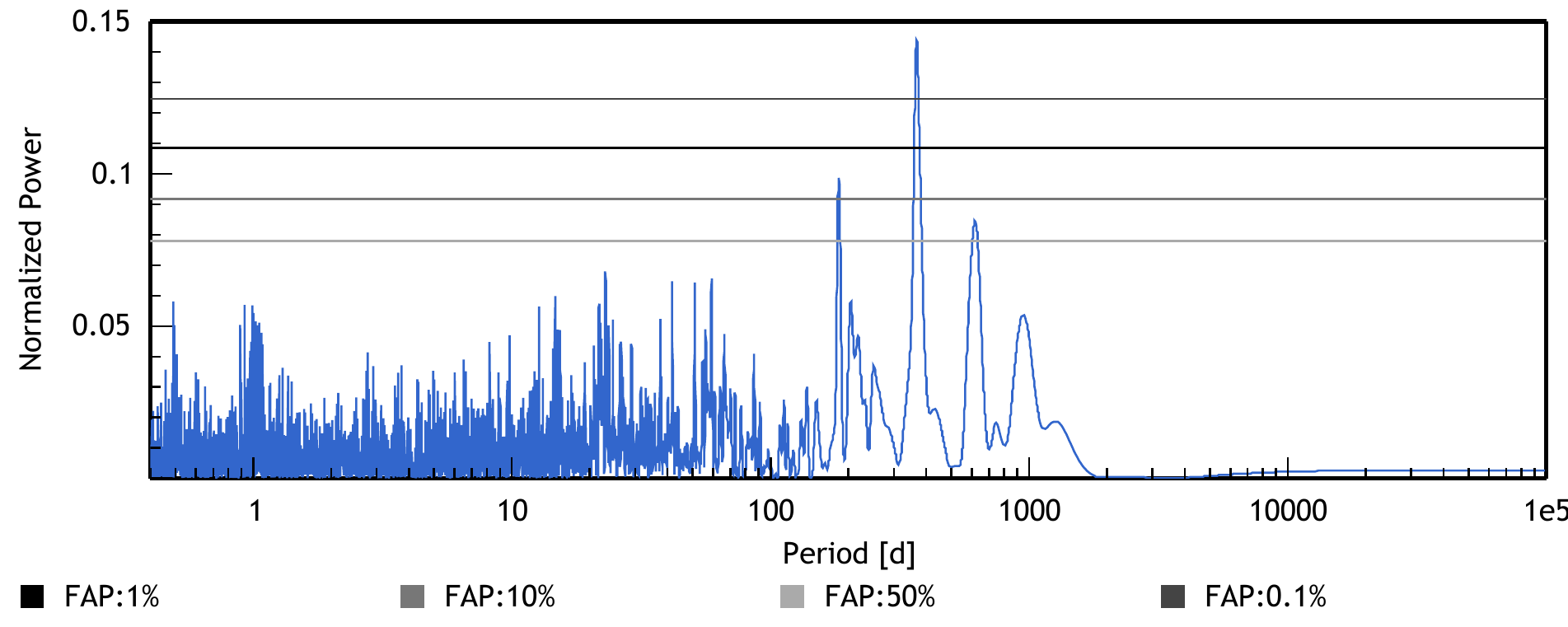}\label{fig:subfig5}}
\hspace*{0.02\textwidth}
\subfloat[HD~1461 new, k3d0.]{\includegraphics[width=0.49\textwidth, keepaspectratio]{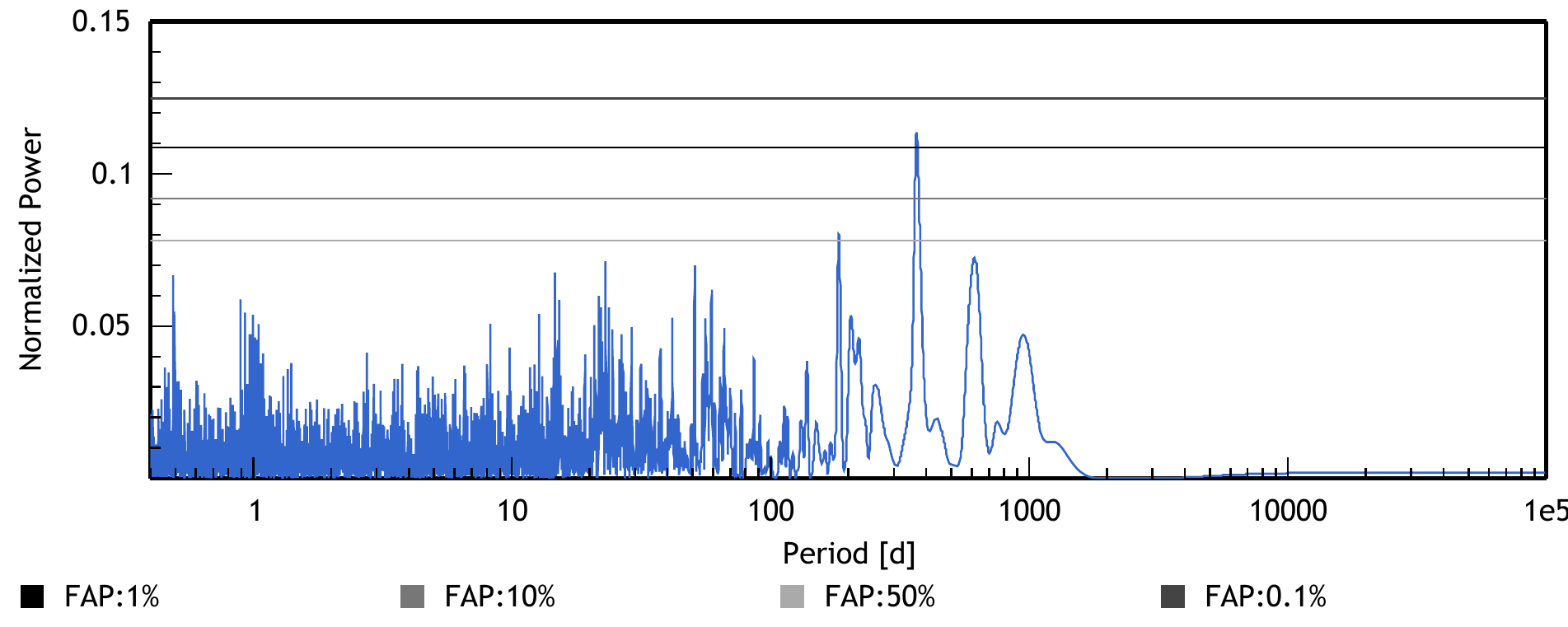}\label{fig:subfig5}}

\caption{Periodograms of HD~1461 after the fit of the two known planets at 5.77 and 13.50~days and the magnetic cycle at 3580~days.}
\label{fig:test}
\end{figure*}


\begin{figure*}
\centering

\subfloat[HD~154088 standard, k1d0.]{\includegraphics[width=0.49\textwidth, keepaspectratio]{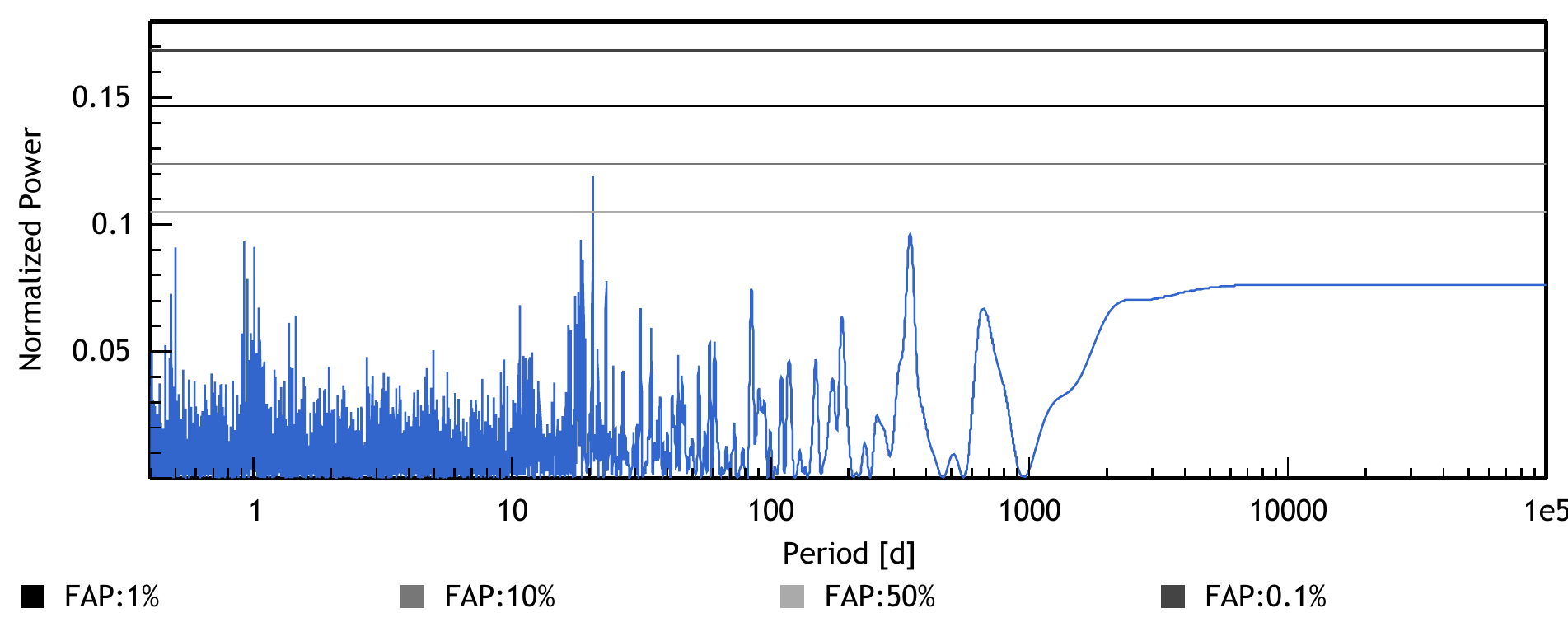}\label{fig:subfig5}}
\hspace*{0.02\textwidth}
\subfloat[HD~154088 new, k1d0.]{\includegraphics[width=0.49\textwidth, keepaspectratio]{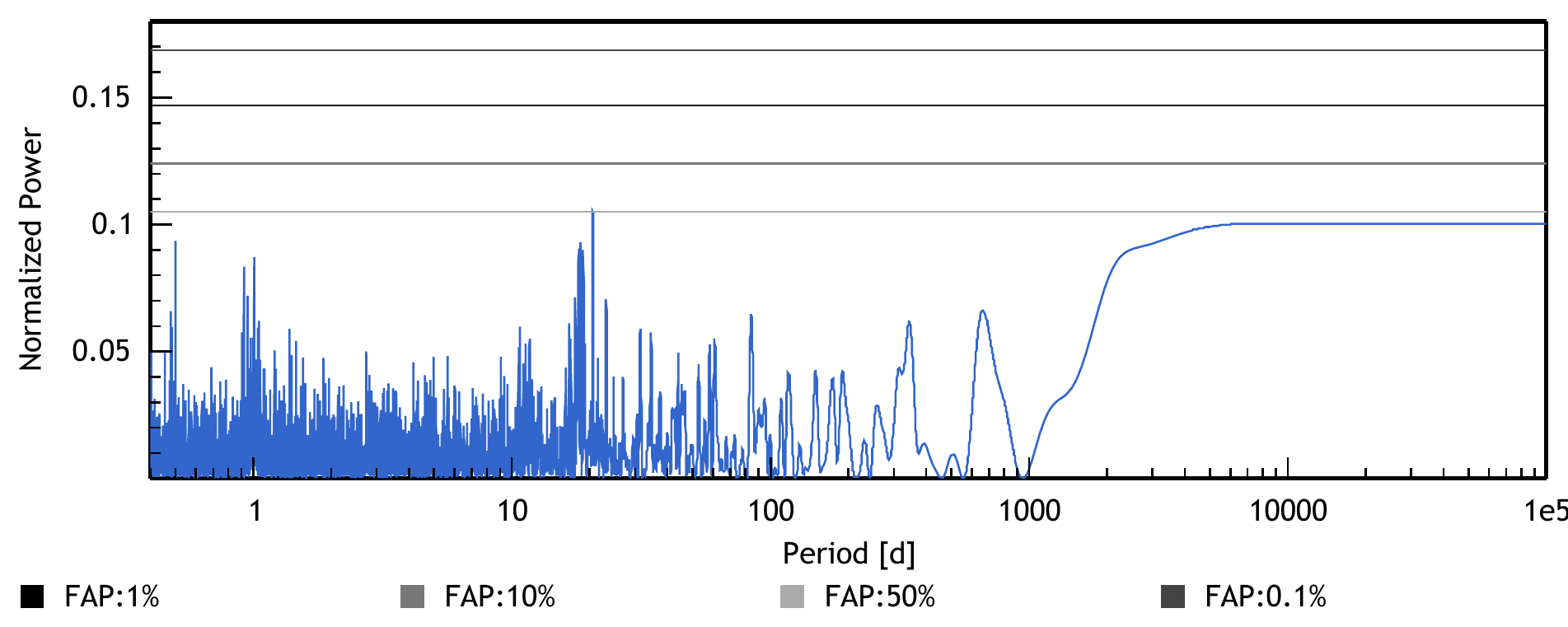}\label{fig:subfig5}}

\caption{Periodograms of HD~154088 after the fit of 1 planet at 18.55 days.}
\label{fig:test}
\end{figure*}


\begin{figure*}
\centering

\subfloat[HD~154088 standard, k1d1.]{\includegraphics[width=0.49\textwidth, keepaspectratio]{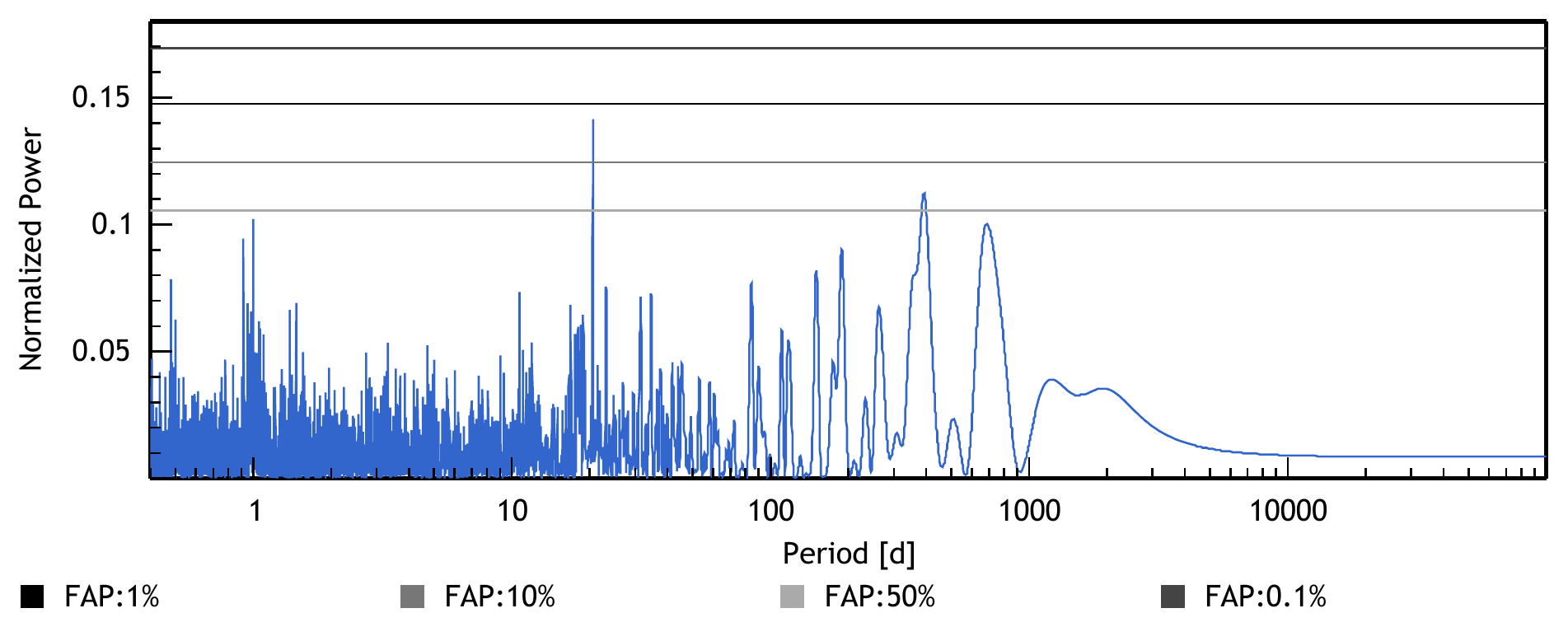}\label{fig:subfig5}}
\hspace*{0.02\textwidth}
\subfloat[HD~154088 new, k1d1.]{\includegraphics[width=0.49\textwidth, keepaspectratio]{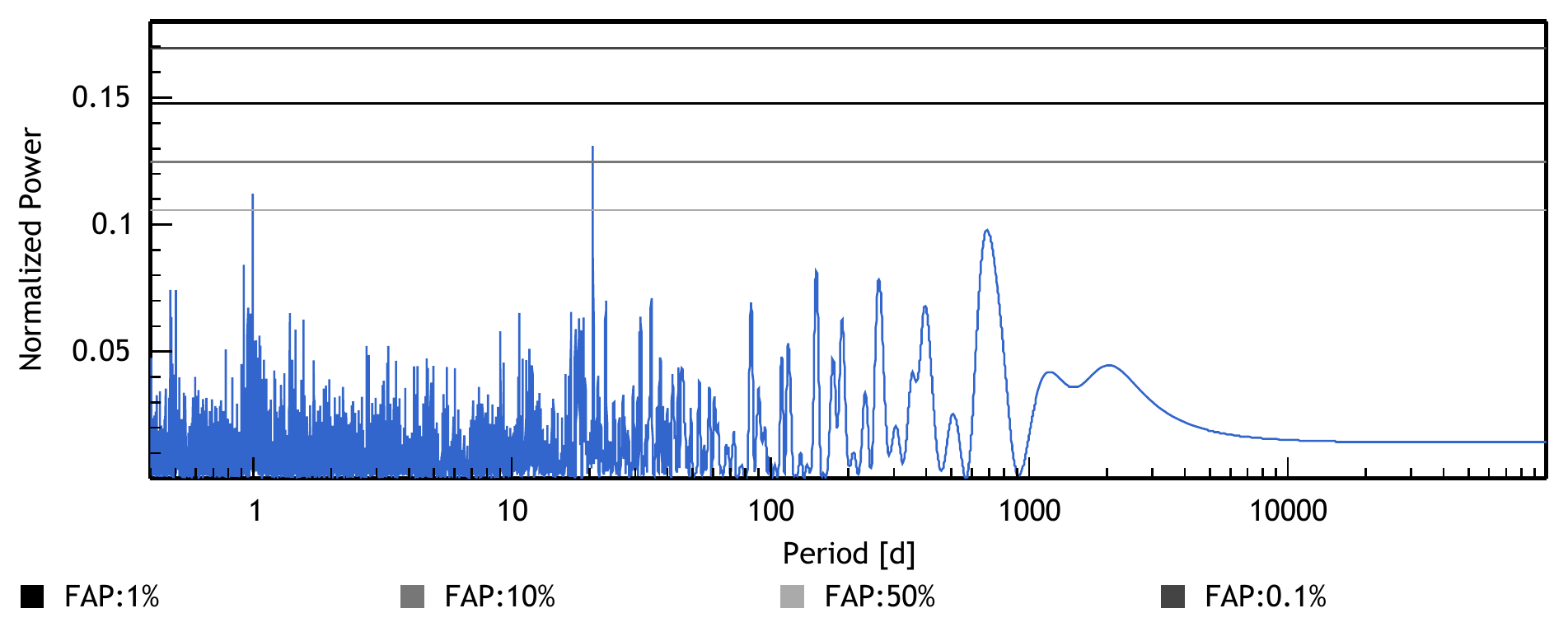}\label{fig:subfig5}}

\caption{Periodograms of HD~154088 after the fit of 1 planet at 18.55 days and a linear drift.}
\label{fig:test}
\end{figure*}


\begin{figure*}
\centering

\subfloat[HD~31527 standard, k3d0.]{\includegraphics[width=0.49\textwidth, keepaspectratio]{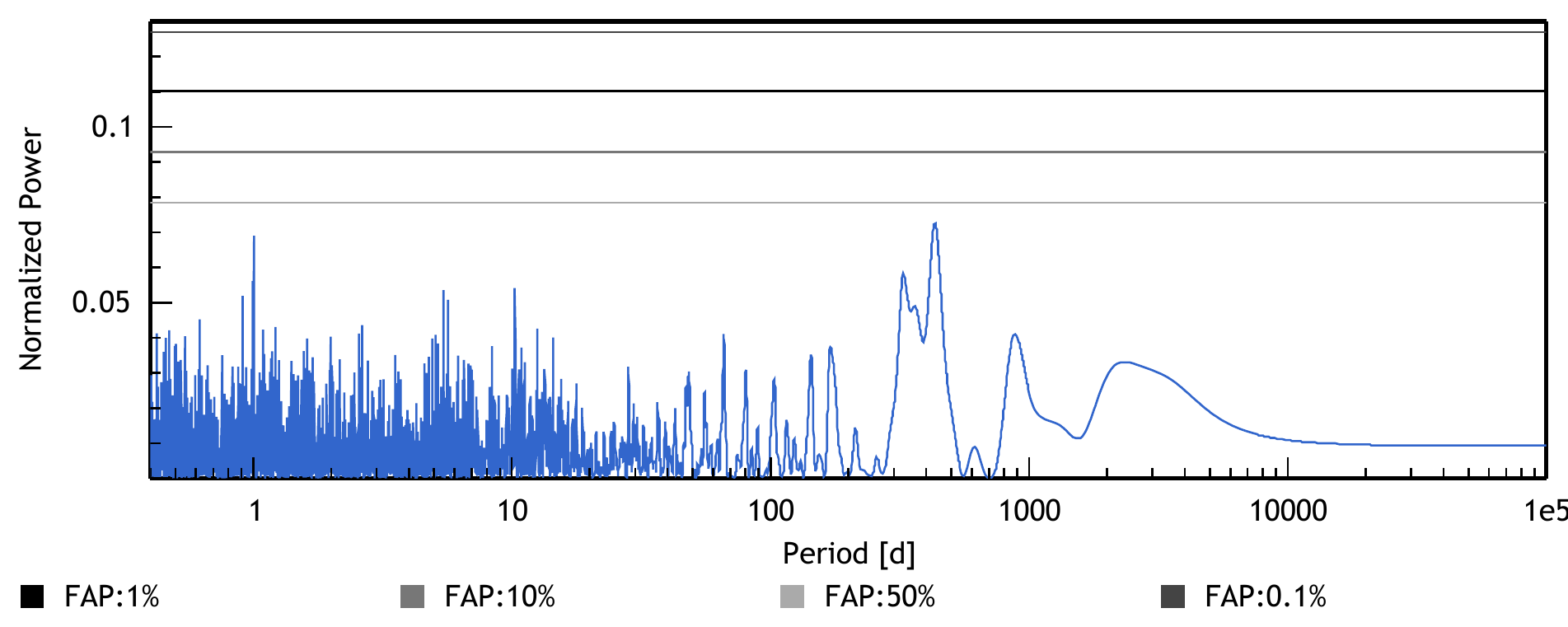}\label{fig:subfig5}}
\hspace*{0.02\textwidth}
\subfloat[HD~31527 new, k3d0.]{\includegraphics[width=0.49\textwidth, keepaspectratio]{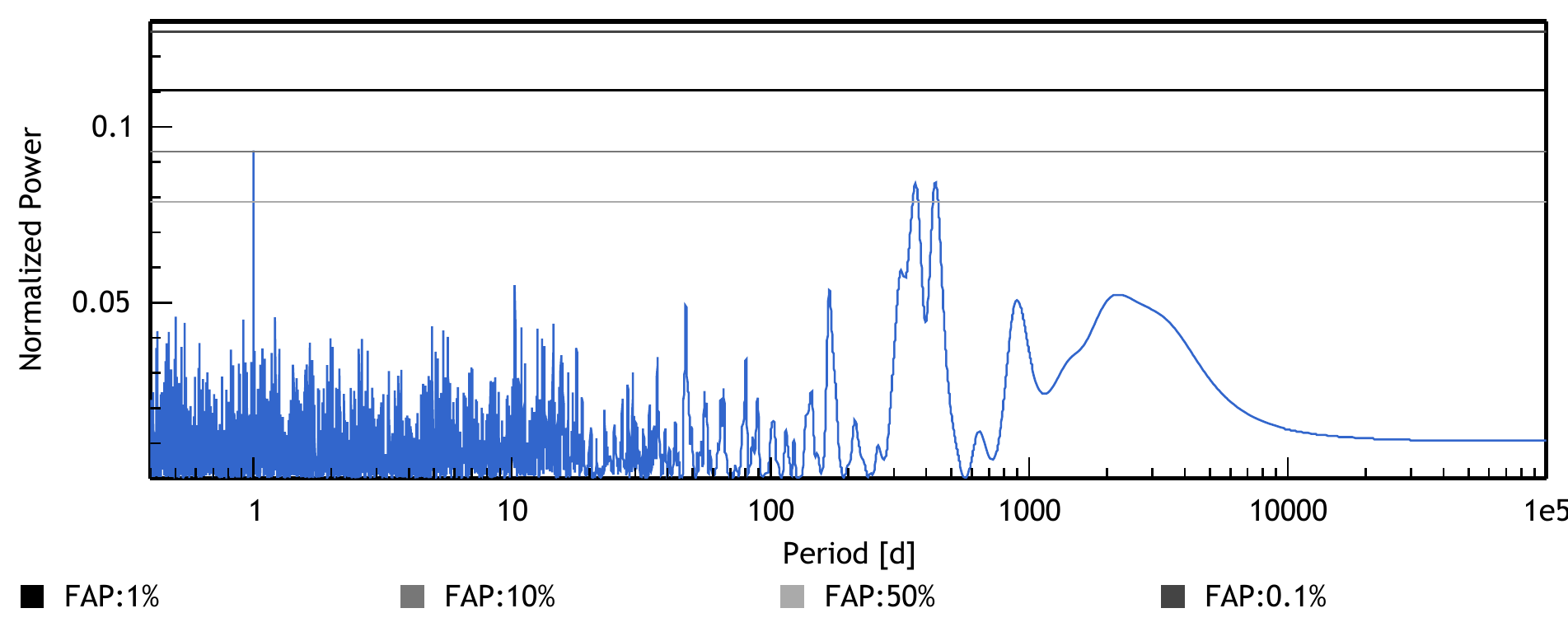}\label{fig:subfig5}}

\caption{Periodograms of HD~31527 after the fit of the 3 known planets: 16.56, 51.2 and 272~days.}
\label{fig:test}
\end{figure*}


\begin{figure*}
\centering

\subfloat[HD~40307 standard, k4d3.]{\includegraphics[width=0.49\textwidth, keepaspectratio]{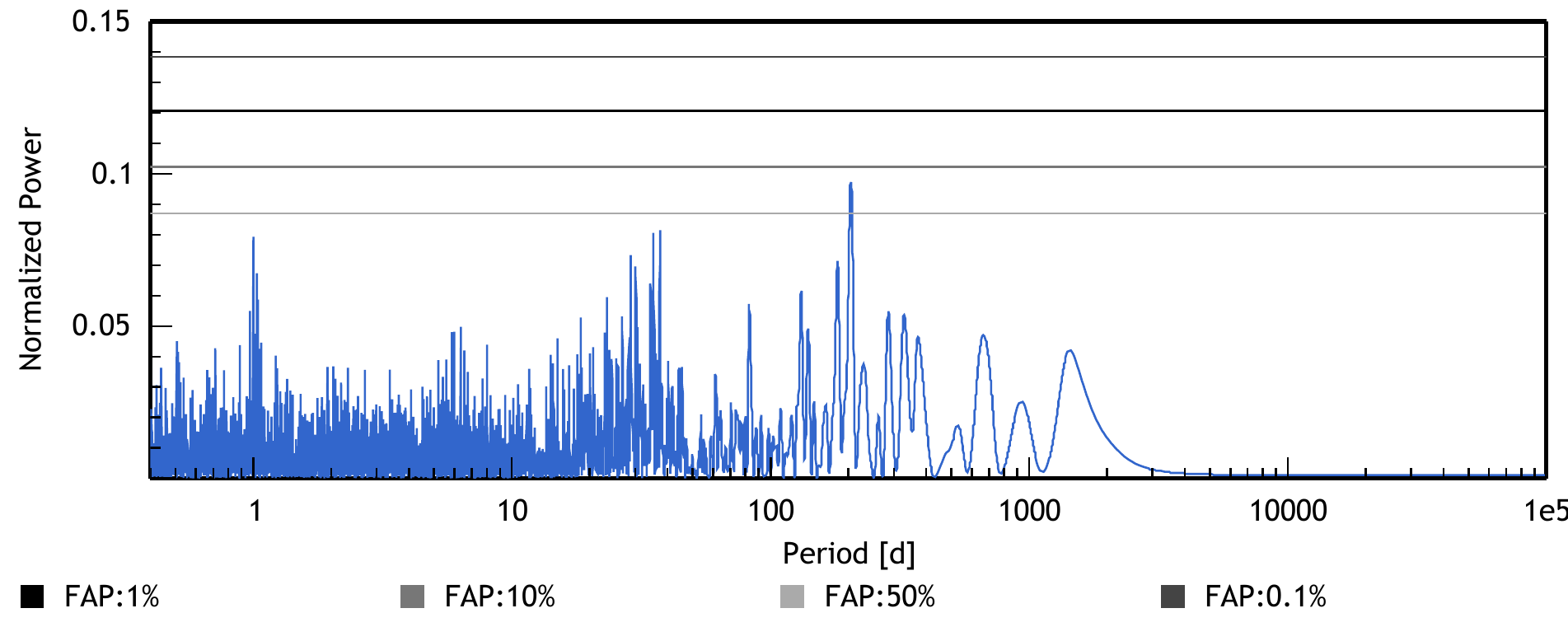}\label{fig:subfig5}}
\hspace*{0.02\textwidth}
\subfloat[HD~40307 new, k4d3.]{\includegraphics[width=0.49\textwidth, keepaspectratio]{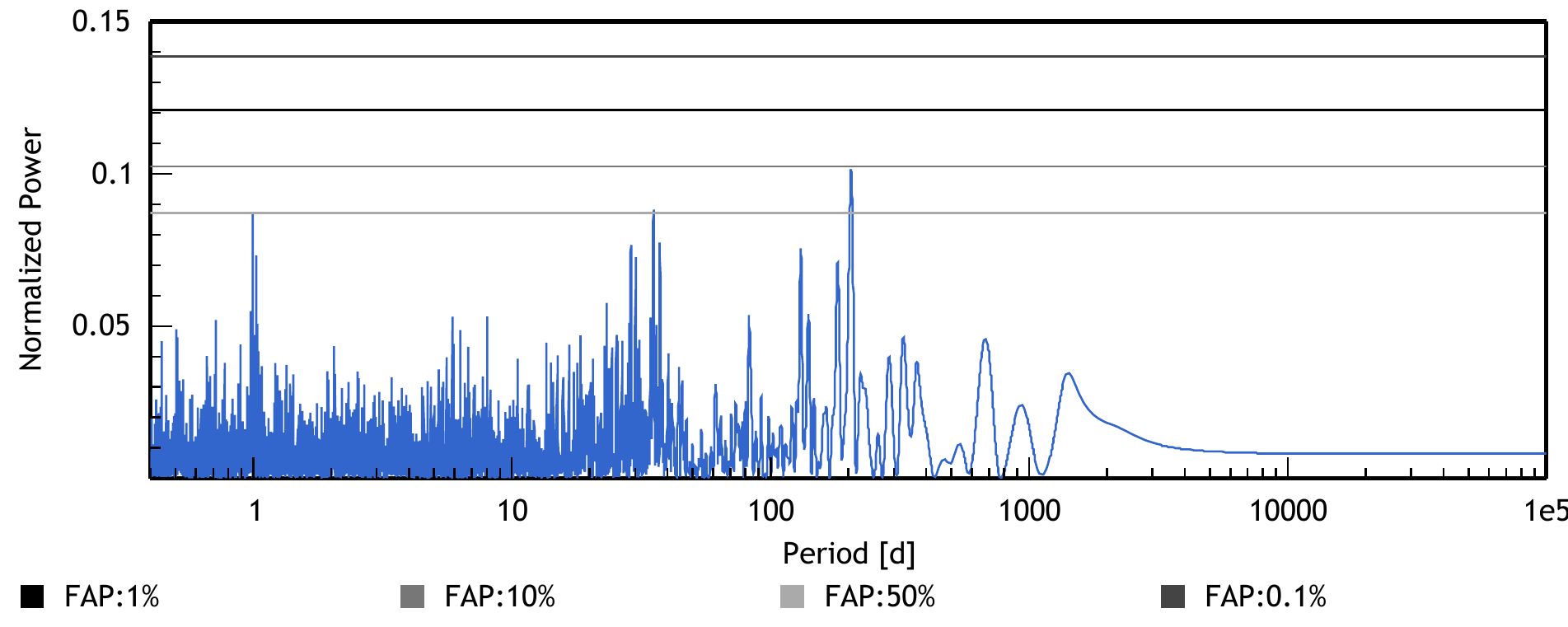}\label{fig:subfig5}}

\caption{Periodograms of HD~40307 after the fit of four signals at periods of 4.31, 9.62, 20.42 and 51.6~days as well as a 3rd-degree polynomial for the magnetic cycle.}
\label{fig:test}
\end{figure*}


\subsubsection{Conclusion}
     \label{subsubsec:Conclusion}

We can see that the global rms is generally slightly smaller with the new version than with the standard version. For the subset of data obtained after the change of fibres, the rms is generally significantly better, as shown by the results of Section~\ref{sec:Global results} and in Figure~\ref{fig:HD220507_postu}.

Concerning the fitted signals (confirmed planets and magnetic cycles), all the amplitudes are compatible within 2~$\sigma$, most of them within 1~$\sigma$, between the two versions of the DRS, as well as between these results and those in the literature.

We also compare the periodogram power around 1~year between the two versions of the DRS. Among the 7~systems studied, the periodogram power at 1~year decreases with the new reduction in 6~cases. In the only case where it slightly increases (HD~31527), the power at 1 year remains below the significance threshold. This demonstrates that spurious instrument-related 1-year signals are significantly reduced with the new DRS. Overall, we note that several systems exhibit a complex periodogram, with significant signals not previously identified as planets at various periods. Through the window function of the data, this excess power will partially leak into other frequencies, in particular at one year, which further complicates our analysis. Nevertheless, among the 7~systems, only one (HD~1461) seems to have a significant signal at 1~year with the new reduction. It is possible that this particular signal is in fact of astrophysical origin.

\subsection{Test on individual stellar spectral lines}
     \label{sec:Test star lines}

\begin{figure*}[htbp!]
   \subfloat[$\lambda=\SI{5031.02}{\AA}$, ordre~39, pixel~3042.]{
      \includegraphics[width=0.45\linewidth]{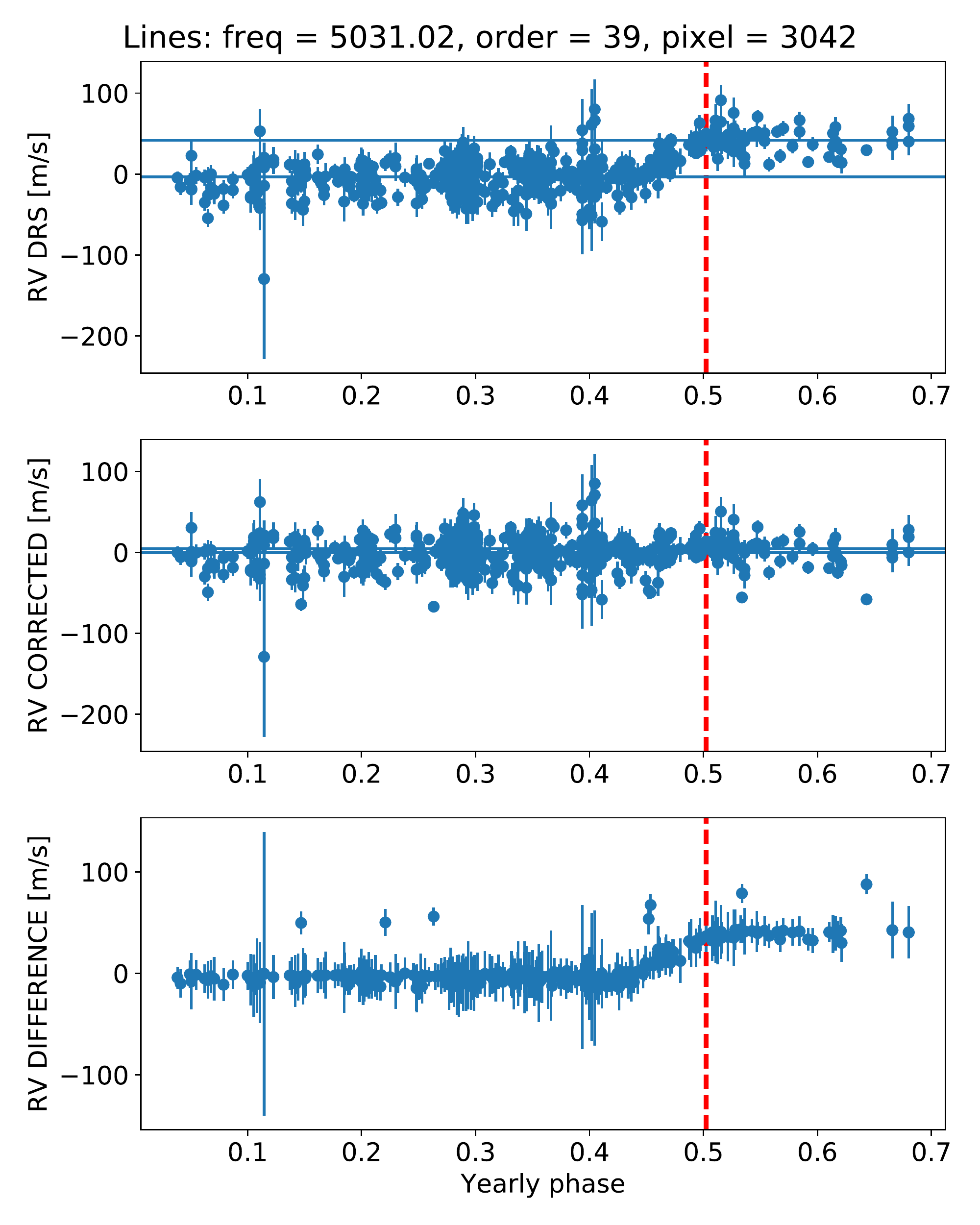}
      \label{fig:Xavier line 1}
   } \quad
   \subfloat[$\lambda=\SI{5393.17}{\AA}$, ordre~47, pixel~504.]{
      \includegraphics[width=0.45\linewidth]{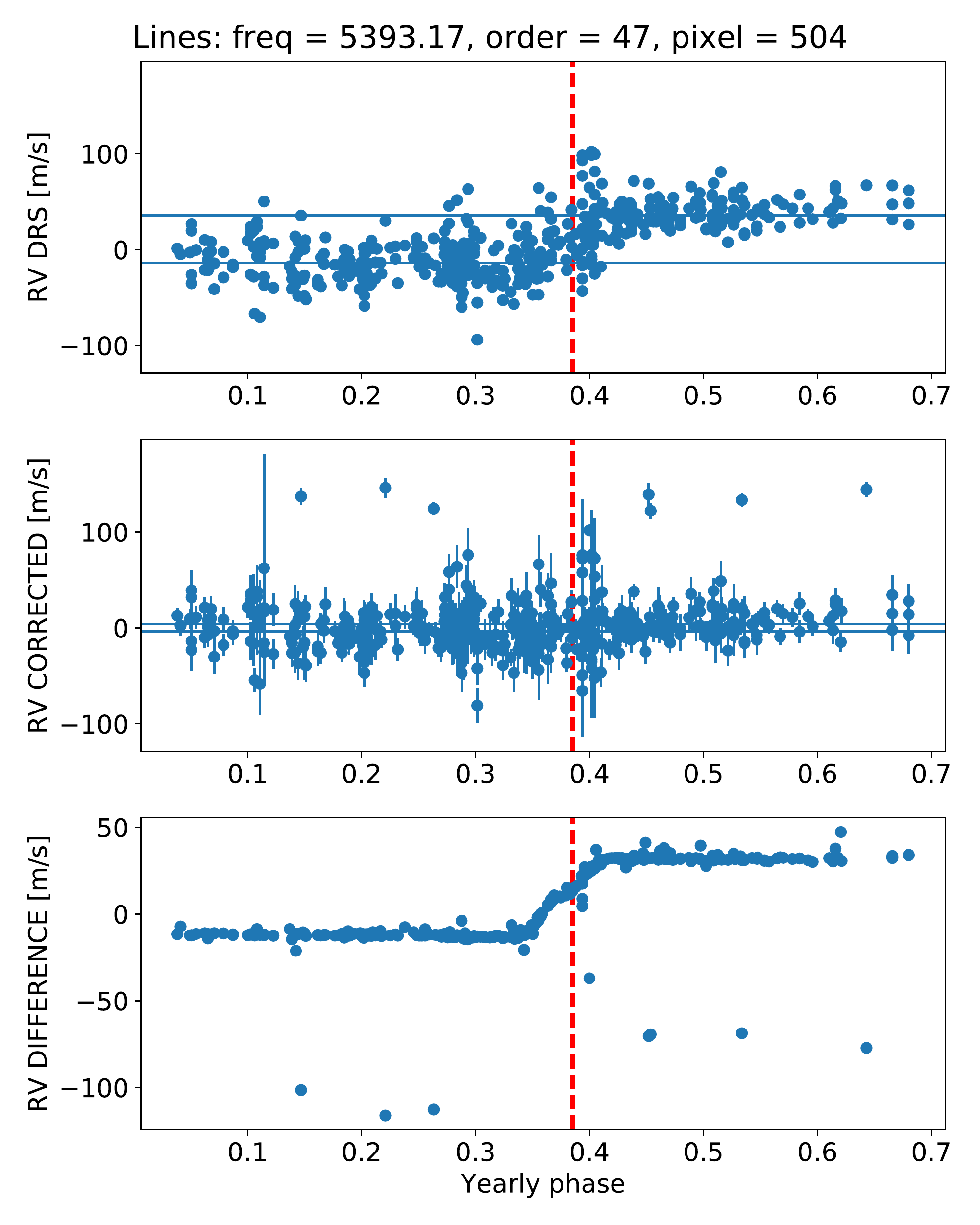}
      \label{fig:Xavier line 2}
   } \quad
   \subfloat[$\lambda=\SI{6393.61}{\AA}$, ordre~64, pixel~3075.]{
      \includegraphics[width=0.45\linewidth]{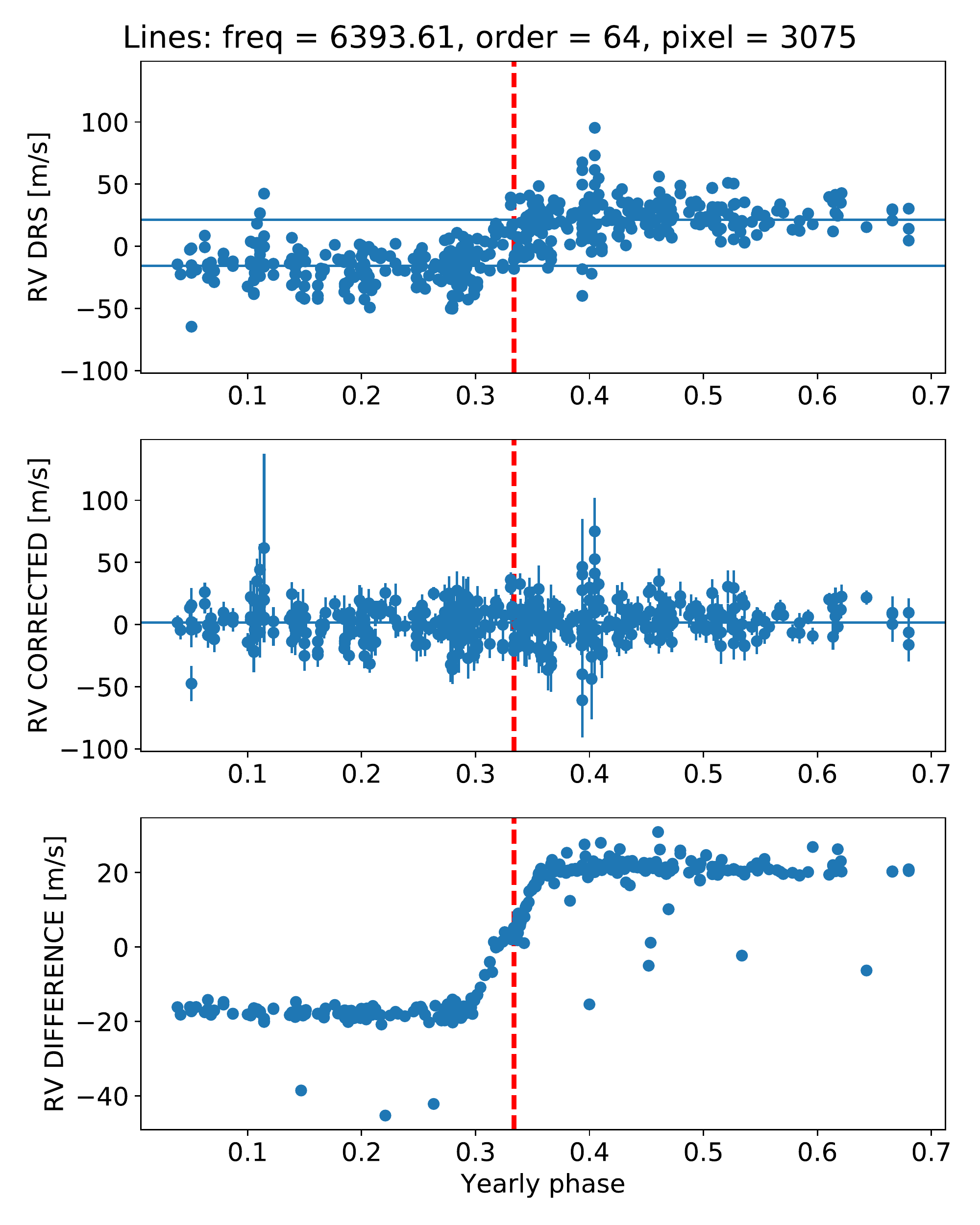}
      \label{fig:Xavier line 1}
   } \\
   \caption{RVs of three spectral lines in the spectrum of HD~1461 that passes over CCD stitching boundaries. We plot those RVs as a function of phase during the year, because due to Earth orbiting around the Sun, the stellar spectrum will move on the CCD with a period of a year. This implies that the spectral lines selected here will pass over CCD stitching boundaries, shown in these plots with the red vertical dashed lines, always at the same phase. The top panels show the RV derived using the standard DRS and the middle panels the RV derived with the correction described in this paper. We show in the bottom panels the difference between those two RV sets. The blue horizontal lines represent the median RV before and after the spectral line passes over the CCD stitching boundary. As we can see, the significant offset between the DRS RVs before and after the CCD stitching boundary disappears in the RVs derived with the correction described in this paper, proving that we consider correctly the different pixel sizes at the stitching positions.}
   \label{fig:Xavier lines}
\end{figure*}

To see the correction that our new wavelength solution brings at the location of the stitchings boundaries, we derived the velocity of a few spectral lines of the star HD~1461 passing over CCD stitching boundaries using the method described in \cite{Dumusque2018}. In Fig.~\ref{fig:Xavier lines}, we show the RV of a few spectral lines passing over CCD stitchings boundaries derived using the standard data reduction software (DRS), in which the wavelength solution is estimated using a unique pixel size, and derived using the formalism described in this paper, in which the wavelength solution is obtained using the real pixel size. We also show in the bottom panels of Fig.~\ref{fig:Xavier lines} the difference between those two sets of RV.

As we can see in the standard DRS RVs, there is a significant drift when a spectral line goes over a CCD stitching boundary, shown in Fig.~\ref{fig:Xavier lines} by red vertical dashed lines. In addition, by looking at the blue horizontal lines that represent the median RV before and after the spectral line passes over the CCD stitching boundaries, we see that these two median RVs are different. This can be explained by the fact that because the pixel at the stitching does not have the same size as the other pixels, the wavelength solution after the line passed over the stitching boundary is under or over-estimated, thus inducing an offset in RV. When looking at the RVs derived using the method described in this paper, the offset between the median RV before and after the stitching boundary disappears, proving that our new wavelength solution considers correctly the different pixel sizes at the stitching positions. This reinforces the idea that the remaining signal around one year for this specific star is not due to this stitching effect but could well be of astrophysical origin.

\section{Conclusion}

   \begin{enumerate}
      \item We manage to estimate the size of the gaps of the CCDs of HARPS that induce a one-year-period parasitic signal observed in the periodogram of several stars. These sizes range from \num{-0.034} to \SI[retain-explicit-plus]{+0.015}{\px}. The validity of this correction was verified with laser frequency comb exposures.
      \item We reprocess all the HARPS wavelength calibration exposures with an improved model taking into account the gaps and the most recent thorium-line wavelengths. We verified that the new wavelength solutions show smaller wavelength distortions than before by comparing to laser-frequency-comb data. Moreover, the new solution are smooth accross bolck stitching boundaries, while the old solutions show significant jumps.
      \item While the RV dispersion of the data obtained before the change of fibres of May~2015 is slightly but consistantly improved, the rms of the data obtained after the change of fibres is significantly better.
      \item A reanalysis of a number of planetary systems with the new pipeline is able to recover all the clearly significant planetary and magnetic signals published in the literature, with fitted parameters fully compatible with the published ones. On the contrary, several weak signals close to the noise level have different properties depending on the pipeline version used.
      \item When we use the new DRS, we systematically reduce the periodogram power at a period of one year in the analysed systems.
      \item The analysis of the RVs derived from individual stellar lines shows that we indeed correct the RV systematic offsets induced by the crossing of stitching boundaries.
   \end{enumerate}

The same wavelength recalibration process can be applied to any other current and future spectrographs, such as the very-high-precision instruments HARPS-N, ESPRESSO, NIRPS, SPIRou and ELT-HIRES.

The next step to further improve the wavelength calibration of fibre-fed spectrographs will consist in the combination of Fabry-Perot and thorium exposures (see e.g.~\citealt{Wildi+2011,SR12,Halverson+2014,Schwab+2015,BZR15,Cersullo+2017}), coupled with the improved procedure described here. This is beyond the scope of the present paper and will be described in details in \cite{CC+2018}.

\begin{acknowledgements}
      We thank the anonymous referee for his/her remarks that helped a lot improve the article. We acknowledge the support of the Swiss National Science Foundation. X.~D. thanks the Branco Weiss Fellowship--Society in Science for its financial support. This publication makes use of the Data \& Analysis Center for Exoplanets (DACE), which is a facility based at the University of Geneva dedicated to extrasolar planets data visualisation, exchange and analysis. DACE is a platform of the Swiss National Centre of Competence in Research (NCCR) PlanetS, federating the Swiss expertise in Exoplanet research. The DACE platform is available at \url{https://dace.unige.ch}. The NCCR are a research instrument of the Swiss National Science Foundation.
      
\end{acknowledgements}

\bibliographystyle{aa}
\bibliography{biblio}

\end{document}